\documentclass[11pt]{article}

\usepackage[preprint]{acl}

\usepackage{times}
\usepackage{latexsym}

\usepackage[T1]{fontenc}

\usepackage[utf8]{inputenc}

\usepackage{microtype}

\usepackage{inconsolata}

\usepackage{graphicx}

\usepackage{CJKutf8}          

\usepackage{amsmath, amsfonts, amssymb}
\usepackage{nicefrac}         
\usepackage{pifont}           

\usepackage{booktabs}         
\usepackage{tabularx}         
\usepackage{multirow}         
\usepackage{makecell}         
\usepackage{siunitx}          
\usepackage{colortbl}         
\usepackage{tablefootnote}    

\usepackage{caption}
\usepackage{wrapfig}          
\usepackage{microtype}        

\usepackage{enumitem}         
\usepackage{tcolorbox}
\tcbuselibrary{skins, breakable} 

\usepackage{algorithm}
\usepackage{algpseudocode}    

\usepackage{underscore}       
\usepackage{verbatim}         
\usepackage{minitoc}          
\usepackage{natbib}           
\usepackage{xcolor}           

\usepackage{url}
\usepackage{hyperref}         
\usepackage{subfiles}         

\newcommand{\mtag}{\textsuperscript{\dag}}

\title{R2MED: A Benchmark for Reasoning-Driven Medical Retrieval}

\author{%
  Xiangxu Zhang$^1$\thanks{Equal contribution.}, Lei Li$^1$\footnotemark[1], Xiao Zhou$^1$\thanks{Xiao Zhou and Zheng Liu are corresponding authors.}, Zheng Liu$^2$\footnotemark[2] \\
  $^1$Gaoling School of Artificial Intelligence, Renmin University of China,\\
  $^2$Beijing Academy of Artificial Intelligence  \\
  \texttt{\{xansar,leil,xiaozhou\}@ruc.edu.cn}, \\
  \texttt{zhengliu1026@gmail.com} \\
}

\begin{document}
\maketitle
\begin{abstract}
  Current medical retrieval benchmarks primarily emphasize lexical or shallow semantic similarity, overlooking the reasoning-intensive demands that are central to clinical decision-making. In practice, physicians often retrieve authoritative medical evidence to support diagnostic hypotheses. Such evidence typically aligns with an inferred diagnosis rather than the surface form of a patient's symptoms, leading to low lexical or semantic overlap between queries and relevant documents. To address this gap, we introduce \textbf{R2MED}, the first benchmark explicitly designed for reasoning-driven medical retrieval. It comprises 876 queries spanning three tasks: Q\&A reference retrieval, clinical evidence retrieval, and clinical case retrieval. These tasks are drawn from five representative medical scenarios and twelve body systems, capturing the complexity and diversity of real-world medical information needs. We evaluate 15 widely-used retrieval systems on R2MED and find that even the best model achieves only 31.4 nDCG@10, demonstrating the benchmark’s difficulty. Classical re-ranking and generation-augmented retrieval methods offer only modest improvements. Although large reasoning models improve performance via intermediate inference generation, the best results still peak at 41.4 nDCG@10. These findings underscore a substantial gap between current retrieval techniques and the reasoning demands of real clinical tasks. We release R2MED as a challenging benchmark to foster the development of next-generation medical retrieval systems with enhanced reasoning capabilities.
\end{abstract}

\section{Introduction}

\begin{figure*}[t]
\includegraphics[width=\linewidth]{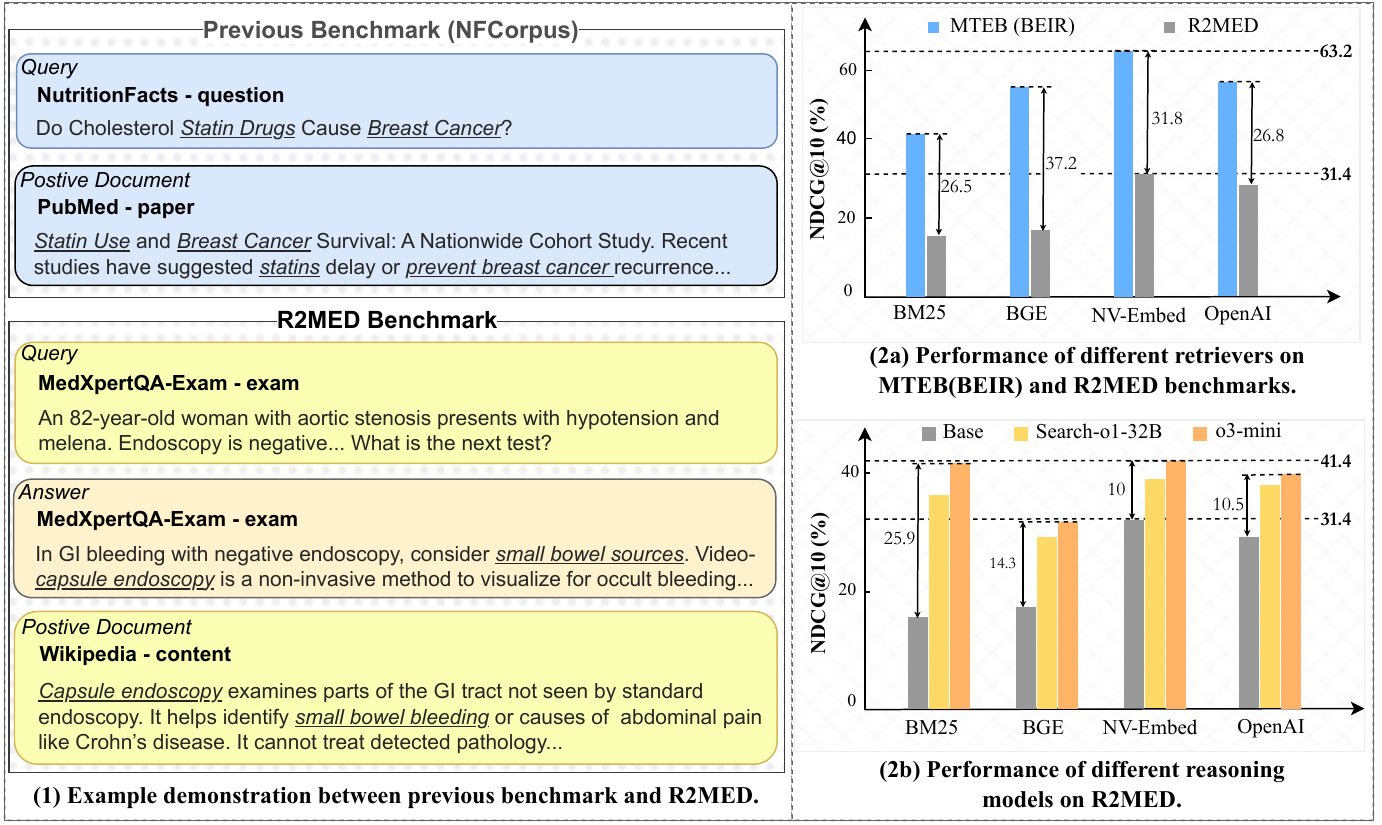}
  \caption{Overview of R2MED. Subfigure (1) compares R2MED with the previous benchmark NFCorpus, illustrating the shift from semantic matching to reasoning-driven retrieval. Subfigures 2(a) and 2(b) show the performance of retrieval and reasoning models on R2MED, highlighting the limitations of existing retrievers on reasoning-driven benchmarks.}
  \label{fig:intro}
\end{figure*}

Medical information retrieval (MIR) plays an important role in clinical decision-making by helping clinicians locate relevant evidence from sources such as biomedical literature, knowledge bases, and clinical records~\cite{luo2008medsearch,goeuriot2016medical,frisoni2022bioreader,zhang2026hypemed}. In real-world settings, however, medical retrieval often requires more than lexical or semantic matching~\cite{lee2019latent,jin2023medcpt}. Relevance may depend on implicit clinical reasoning, such as inferring latent symptom--disease associations or identifying evidence that supports a diagnostic conclusion not explicitly stated in the query. Existing MIR benchmarks~\cite{boteva2016full,voorhees2021trec,li2024automir} largely do not capture this challenge. For example, benchmarks such as NFCorpus~\cite{boteva2016full} align queries and documents through explicit links, so lexical overlap becomes a strong signal of relevance (Figure~\ref{fig:intro} (1)). As a result, these benchmarks mainly reward shallow matching and fail to reflect the reasoning-intensive retrieval needs of clinical practice.

This limitation has become more important in modern medical question answering~\cite{jin2021disease,zuo2025medxpertqa,qiu2025quantifying}. Retrieval-augmented generation (RAG) depends on retrieving evidence that genuinely supports answer generation, while large reasoning models (LRMs) place increasing emphasis on multi-step clinical reasoning~\cite{xiong2024benchmarking,wu2024medical,xiong2024improving,jaech2024openai,chen2024huatuogpt,zhang2025inf}. In both settings, useful evidence is often related to the query through implicit reasoning rather than direct surface overlap. This creates a clear need for benchmarks that explicitly evaluate reasoning-intensive medical retrieval.

In this work, we introduce R2MED, the first benchmark explicitly designed for reasoning-intensive medical retrieval. Unlike prior datasets that emphasize direct lexical overlap or shallow semantic similarity, R2MED evaluates whether a system can retrieve documents that follow the implicit reasoning path needed to answer a query (Figure~\ref{fig:intro} (1)). R2MED contains 876 queries spanning three reasoning-centric retrieval tasks and eight datasets, covering diverse clinical scenarios. Across these tasks, relevance is defined by whether a document supports the inferred diagnosis, conclusion, or clinical reasoning path behind the query, rather than by explicit query--document matching alone.

We benchmark 15 classical retrieval systems on R2MED and observe a substantial drop in performance relative to standard retrieval benchmarks. For example, although NV-Embed-v2~\cite{lee2024nv} reaches 63.2 nDCG@10 on the MTEB~\cite{muennighoff2022mteb} retrieval subset of BEIR~\cite{thakur2021beir}, it achieves only 31.4 nDCG@10 on R2MED (Figure~\ref{fig:intro} (2a)). We also study reasoning-augmented retrieval with LRMs and find that reasoning can improve performance on complex medical queries, but only to a limited degree. Even the strongest configuration, NV-Embed-v2 augmented with o3-mini reasoning guidance, reaches only 41.4 nDCG@10 (Figure~\ref{fig:intro} (2b)). These results indicate that reasoning-intensive medical retrieval remains an open challenge.

\noindent \textbf{Contributions}: {(1)} we introduce R2MED, the first benchmark for reasoning-intensive medical retrieval across three tasks and eight datasets; {(2)} we provide a comprehensive evaluation of 15 retrieval systems, showing that strong retrievers still perform poorly on R2MED, with the best nDCG@10 only reaching 31.4; and {(3)} we analyze reasoning-augmented retrieval with large reasoning models, showing that although reasoning improves retrieval, current methods remain insufficient for the demands of complex medical retrieval scenarios.

\section{Related Work}
\noindent \textbf{Medical Retrieval Benchmarks.} 
To support the advancement of medical information retrieval, a range of domain-specific benchmarks have been developed. Most existing benchmarks like NFCorpus~\cite{boteva2016full}, SciFact~\cite{wadden2020fact}, TREC-COVID~\cite{voorhees2021trec}, and CMIRB~\cite{li2024automir} primarily focus on keywords or shallow semantic matching between the query and relevant documents. For instance, NFCorpus aligns layperson health questions with scientific articles from NutritionFacts.org, using curated links to PubMed literature to establish relevance. Closest to our work, BRIGHT~\cite{su2024bright} begins to explore reasoning-based retrieval by constructing a large-scale dataset of user queries paired with relevant web documents, primarily sourced from community QA forums. However, we take a different perspective by constructing retrieval tasks grounded in authentic clinical scenarios that inherently require multi-step medical reasoning. R2MED is a benchmark dedicated to reasoning-centric medical retrieval, in which relevant documents are often connected to queries through complex reasoning.

\noindent \textbf{Medical QA Benchmarks.} 
Early medical QA benchmarks such as MedQA~\cite{jin2021disease}, MedMCQA~\cite{pal2022medmcqa}, and MMLU (Medical)~\cite{hendrycks2020measuring} are primarily derived from medical licensing and entrance examinations. These benchmarks focus on basic medical knowledge understanding in standardized, multiple-choice formats. Recently, some work has shifted focus toward clinical reasoning and complex medical QA. MedXpertQA~\cite{zuo2025medxpertqa} presents complex, specialty-specific multiple-choice questions grounded in real clinical settings. MedRBench~\cite{qiu2025quantifying} constructs open-ended diagnostic and therapeutic reasoning tasks derived from curated patient case reports. These benchmarks reflect a growing emphasis on robust, multi-step clinical reasoning in medical QA. We curate retrieval-focused queries from a subset of these complex QA datasets and enhance them with additional annotations to construct R2MED.


\noindent \textbf{Dense Retrieval.}
Modern information retrieval has evolved significantly with the rise of dense retrieval models, which encode queries and documents into continuous vector spaces. Representative models such as Contriever~\cite{izacard2021unsupervised}, BGE~\cite{xiao2024c}, BMRetriever~\cite{xu2024bmretriever}, and GritLM~\cite{muennighoff2024generative} are typically pre-trained on large-scale corpora and further fine-tuned using supervised or synthetic data. Generation-augmented retrieval methods such as HyDE~\cite{gao2022precise} and Query2doc~\cite{wang2023query2doc} further reduce the semantic gap by synthesizing hypothetical documents. More recent work incorporates explicit reasoning into retrieval, such as inference-time logical reasoning~\cite{faltings2025enhancing} and logic-aware multi-hop retrieval frameworks like HopRAG~\cite{liu2025hoprag}. In parallel, large reasoning models such as o1~\cite{jaech2024openai} and DeepSeek-R1~\cite{guo2025deepseek}, as well as agentic search systems like Search-o1~\cite{li2025search} and Search-r1~\cite{jin2025search}, have shown promise on complex reasoning-intensive retrieval tasks. In this work, we evaluate these emerging paradigms under a unified reasoning-centric medical retrieval setting, revealing their strengths and limitations.

\section{R2MED: A New Reasoning-Driven Retrieval Benchmark}

\begin{table*}[!t]
  \centering
  \resizebox{\textwidth}{!}{
  \begin{tabular}{l|ccccc|c|c|l}
    \toprule
    Dataset & \multicolumn{1}{c}{\#Q} & \multicolumn{1}{c}{\#D} & \multicolumn{1}{c}{Avg. Pos} & \multicolumn{1}{c}{Q-Len} & D-Len &Q-Source & D-Source &Example \\
\midrule
\rowcolor{gray!8} \multicolumn{9}{l}{\textbf{Q\&A Reference Retrieval Task}} \\
\midrule
Biology &103 &57,359 &3.6 &115.2 &83.6  &\multirow{3}{9em}{\centering StackExchange post~\cite{stackexchange_platform}} & \multirow{3}{9em}{\centering Web pages: article, blog, wikipedia ...} & Tab.~\ref{tab:example_biology} \\
Bioinformatics &77 &47,473 &2.9 &273.8 &150.5 & & \multicolumn{1}{c|}{} & Tab.~\ref{tab:example_bioin} \\
Medical Sciences &88 &34,810 &2.8 &107.1 &122.7 & \multicolumn{1}{c|}{}  & \multicolumn{1}{c|}{} & Tab.~\ref{tab:example_medical} \\
\midrule
\rowcolor{gray!8} \multicolumn{9}{l}{\textbf{Clinical Evidence Retrieval Task}} \\
\midrule
MedXpertQA-Exam &97 &61,379 &3.0 &233.2 &154.9   &{\centering Exam question~\cite{zuo2025medxpertqa}} & {\centering Wikipedia~\cite{xiong2024benchmarking}} & Tab.~\ref{tab:example_medxpert} \\
MedQA-Diag &118 &56,250 &4.4 &167.8 &179.7 &{\centering Exam question~\cite{jin2021disease}} & {\centering Textbooks~\cite{jin2021disease}} & Tab.~\ref{tab:example_medqa} \\
PMC-Treatment &150 &28,954 &2.1 &449.3 &149.3 &{\centering Clinical question~\cite{qiu2025quantifying}} & {\centering PubMed articles~\cite{qiu2025quantifying}} & Tab.~\ref{tab:example_pmct} \\
\midrule
\rowcolor{gray!8} \multicolumn{9}{l}{\textbf{Clinical Case Retrieval Task}} \\
\midrule
PMC-Clinical &114 &60,406 &2.2 &182.8 &480.4  &{\centering Clinical Case~\cite{zhao2023large}} & {\centering PubMed cases~\cite{zhao2023large}} & Tab.~\ref{tab:example_pmcc} \\
IIYi-Clinical &129 &10,449 &3.5 &602.3 &1,273.0 &{\centering Clinical Case~\cite{iiyi_online}} & {\centering IIYi cases} & Tab.~\ref{tab:example_iiyi} \\
\bottomrule
  \end{tabular}
  }
    \caption{Statistics of R2MED. \#Q and \#D denote the number of queries and documents, respectively. Avg. Pos refers to the average positive documents per query. Q-Len and D-Len are the average lengths of queries and documents. We measure the average length by the GPT-2 tokenizer~\cite{radford2019language}.}
\label{tab:statistics}
\end{table*}

\subsection{Preliminary}
Reasoning-driven medical information retrieval poses unique challenges that go beyond surface-level lexical or semantic matching. Formally, given a query $q$ and a document corpus $\mathcal{D} = \{d_1, \ldots, d_n\}$, the task is to identify a subset of relevant documents $\mathcal{D}^+_q = \{D_{q,1}^+, \ldots, D_{q,m}^+\} \subset \mathcal{D}$, where $m \ll n$. All remaining documents are treated as negative examples, denoted by $\mathcal{D}_q^- = \mathcal{D} \setminus \mathcal{D}_q^+$. Unlike conventional retrieval tasks, relevance in this context is mediated by a latent reasoning answer $\mathcal{A}$ that logically links the query to its corresponding positive documents. Importantly, this reasoning answer is often absent from the query’s surface form, requiring models to infer it implicitly via reasoning.

\subsection{Task Curation}
R2MED is a benchmarking dataset designed to evaluate retrieval systems in reasoning-intensive medical scenarios. It comprises three tasks: Q\&A reference retrieval, clinical evidence retrieval, and clinical case retrieval, each targeting a distinct type of clinical information need (see Table~\ref{tab:statistics}).

The Q\&A reference retrieval task aims to retrieve high-quality external resources that provide essential evidence for answering medical questions. Each query is a natural language question sourced from a community post on the StackExchange platform. Relevant documents refer to webpages cited within the corresponding answer, having undergone expert validation to ensure they convey critical knowledge essential for answering the question. Therefore, the answer serves as an implicit reasoning anchor that links the question to its relevant documents.

The clinical evidence retrieval task focuses on retrieving medical evidence that supports diagnostic or treatment planning within the clinical decision-making scenario. Each query is a complex clinical question drawn from established medical QA datasets. Relevant documents are curated from authoritative medical encyclopedias and verified to provide sufficient evidence for the clinical decision implied by the query. The original answer in the QA dataset thus serves as an implicit reasoning step that bridges the query and its relevant documents.

The clinical case retrieval task centers on retrieving similar cases with the same diagnosis to assist in analyzing a given patient scenario. Each query is a structured clinical description, including chief complaint, history, and physical findings, sourced from case reports or electronic health records. Relevant documents are clinical cases sharing the same diagnosis and verified to provide informative support for the query. Here, the diagnosis serves as a latent reasoning bridge linking the query to its relevant documents.

\subsection{Benchmark Construction}

\noindent {\textbf{Data Collection.}}
\label{sec_benchmark_construction}

\begin{figure}[htbp]
\centering
\includegraphics[width=0.9\linewidth]{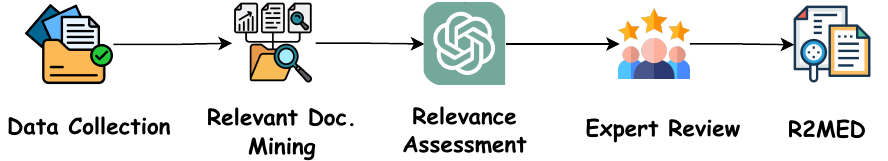}
  \caption{R2MED benchmark construction pipeline.}
  \label{fig:data_pipeline}
\end{figure}
As illustrated in Figure~\ref{fig:data_pipeline}, our dataset construction begins with a systematic and task-specific collection process grounded in high-quality medical corpora. R2MED comprises eight datasets drawn from diverse sources, reflecting variations in data modalities (Table~\ref{tab:statistics}). At this stage, we curate a unified quadruple ($\mathcal{Q}$, $\mathcal{A}$, $\mathcal{D}_\mathrm{init}^+$, $\mathcal{D}_\mathrm{init}^-$) for each dataset, representing the query, gold answer, initial positive documents, and initial negative documents, respectively. For the Q\&A reference retrieval task, we curate query–answer pairs ($\mathcal{Q}$, $\mathcal{A}$) from three StackExchange communities, namely Biology, Bioinformatics, and Medical Sciences, by selecting posts with accepted or highly upvoted answers. The webpages linked within these answers form $\mathcal{D}_\mathrm{init}^+$, while negative documents $\mathcal{D}_\mathrm{init}^-$ are sampled from Wikipedia. Among these, the Biology dataset is adopted directly from the BRIGHT benchmark~\cite{su2024bright}. For the clinical evidence retrieval task, we reformat three medical QA datasets (MedXpertQA~\cite{zuo2025medxpertqa}, MedQA~\cite{jin2021disease}, MedRBench~\cite{qiu2025quantifying}) into three specialized datasets, each corresponding to a different stage of clinical decision-making: examination recommendation, diagnosis, and treatment planning. Candidate documents $\mathcal{D}$ are drawn from three high-quality sources: Wikipedia~\cite{xiong2024benchmarking}, medical textbooks~\cite{jin2021disease}, and PubMed articles~\cite{qiu2025quantifying}. For the clinical case retrieval task, we collect full patient records from PMC-Patients~\cite{zhao2023large} and the IIYi-bingli website. We extract the structured clinical presentation as $\mathcal{Q}$, and the confirmed diagnosis as $\mathcal{A}$ from each record by GPT-4o\footnote{GPT-4o refers to gpt-4o-2024-11-20 in this work.}. Clinical cases with the same diagnosis form $\mathcal{D}_\mathrm{init}^+$, while other cases form $\mathcal{D}_\mathrm{init}^-$. We also apply a series of filtering and restructuring steps to ensure that the resulting queries align with the intended retrieval tasks. Please refer to Appendix~\ref{app:data_collection} for further details.

\noindent {\textbf{Relevant Document Mining.}}
While each dataset in R2MED is initially constructed with a quadruple ($\mathcal{Q}$, $\mathcal{A}$, $\mathcal{D}_{\mathrm{init}}^+$, $\mathcal{D}_{\mathrm{init}}^-$), the negative set $\mathcal{D}_{\mathrm{init}}^-$ may contain false negatives that are relevant but unverified~\cite{chen2024air,moreira2024nv}. To enrich the positive document pool and mitigate noise in negatives, we adopt a retrieval-based mining strategy. Specifically, for each pair ($q$, $a$), we use OpenAI o3 model to generate a step-by-step reasoning path $s$, forming a multi-view retrieval set $\mathcal{S}_q=\{q,a,s\}$. To ensure retrieval diversity, we employ a retrieval committee $\mathcal{C}=\{r_1,r_2,...r_n\}$, where each $r_i$ denotes a distinct retriever. For each element in $\mathcal{S}_q$, each $r_i\in \mathcal{C}$ independently retrieves top-$k$ documents from $\mathcal{D}_\mathrm{init}^-$. We aggregate the retrieved results from all committee members and rank candidate documents based on their frequency of appearance. The top-$k$ most frequently retrieved documents are selected as the mined relevant set $\mathcal{D}_{q,\mathrm{ret}}$, which is merged with the initial positives to form the enhanced positive pool $\mathcal{D}_{q,\mathrm{pool}}^+$. Simultaneously, these documents are removed from $\mathcal{D}_{q,\mathrm{init}}^-$ to update the negative pool $\mathcal{D}_{q,\mathrm{pool}}^-$. This process yields an intermediate quadruple ($\mathcal{Q}, \mathcal{A}, \mathcal{D}_{\mathrm{pool}}^+, \mathcal{D}_{\mathrm{pool}}^-$). See Appendix~\ref{app:relevant_mining} for more details.

\noindent {\textbf{Relevance Assessment.}}
To ensure data quality, we perform a fine-grained relevance assessment on the pooled document sets using GPT-4o. For each candidate document $d \in \mathcal{D}_{q,\mathrm{pool}}^+$, we evaluate its relevance using the triple $(q, a, s)$. The assessment follows a two-dimensional scoring rubric on a 0–10 scale, assessing i) the document's relevance to the answer, ii) its support for the reasoning process. Documents scoring at least 8 in both dimensions are retained as verified positives $\mathcal{D}_{q,\mathrm{ver}}^+$. Those receiving ambiguous scores (5–7) are discarded to avoid introducing noise into the evaluation. Documents scoring 4 or below are treated as verified negatives and added to the set $\mathcal{D}_{q,\mathrm{ver}}^-)$. This procedure yields a refined and rigorously validated dataset $(\mathcal{Q}, \mathcal{A}, \mathcal{D}_{\mathrm{ver}}^+, \mathcal{D}_{\mathrm{ver}}^-)$. Full details of the scoring protocol are provided in Appendix~\ref{app:relevance_assessment}.

\noindent {\textbf{Expert Review.}}
To ensure clinical validity and factual reliability, especially in light of the involvement of language models in data generation and assessment, we conduct a final expert review stage. In this stage, a medically trained reviewer examines all data samples to identify potential quality issues. A board-certified medical expert then re-examines the flagged cases to make the final judgment. Each data point is reviewed across three criteria: (1) whether the reformulated query (if applicable) is clinically coherent and complete; (2) whether the reasoning path reflects plausible and accurate medical inference; and (3) whether the positive documents provide essential support for both the query and answer. Data that fail to meet these criteria are excluded from the final release. Additional details are provided in Appendix~\ref{app:expert_review}.

\subsection{Diversity Analysis}
We assess the diversity of R2MED from both clinical and distributional perspectives. Each query is categorized by its medical scenario and involved body system. As shown in Figure~\ref{fig:bench_att}, R2MED covers 5 major clinical scenarios and 12 body systems, capturing a wide range of real-world medical contexts. In addition, we compute weighted Jaccard similarity across datasets and observe consistently low overlap, indicating that R2MED presents a challenging testbed requiring strong generalization across diverse and out-of-distribution domains. See Appendix~\ref{app:data_diversity_analysis} for more details.

\section{Experiments}
\label{sec:experiments}

\subsection{Experimental Setup}
We evaluate 15 representative retrieval models, including both sparse retrieval (BM25~\cite{robertson2009probabilistic}) and dense retrieval models (top performers on the MTEB leaderboard~\cite{mteb_leaderboard}). Dense retrieval models are divided into two categories: base-size models (< 1B) such as Contriever~\cite{izacard2021unsupervised}, MedCPT~\cite{jin2023medcpt}, InstructOR-L~\cite{su2022one}, BGE-Large\cite{xiao2024c}, and BMRetriever-410M~\cite{xu2024bmretriever}, and large-size models (> 1B) including InstructOR-XL~\cite{su2022one}, BMRetriever-2B/7B~\cite{xu2024bmretriever}, E5-Mistral~\cite{wang2023improving}, GritLM-7B~\cite{muennighoff2024generative}, SFR-Embedding-Mistral~\cite{meng2024sfrembedding}, NV-Embed~\cite{lee2024nv}. We additionally evaluate two proprietary embedding models from OpenAI~\cite{openaiemb} and Voyage~\cite{voyageemb}. Among these, MedCPT and the BMRetriever family are domain-specific retrievers pretrained on large-scale biomedical corpora. Detailed model descriptions are provided in Appendix~\ref{app:model_details}. Following prior work~\cite{nguyen2016ms, thakur2021beir, su2024bright}, we use nDCG@10 as the primary evaluation metric.

\subsection{Main Results}
\begin{table*}[t!]
\setlength{\tabcolsep}{3pt}

\centering
\resizebox{\textwidth}{!}{
\begin{tabular}{l|c|ccc|ccc|cc|c}
\toprule 
Task &\multirow{2}{*}{\centering Size} & \multicolumn{3}{c|}{Q\&A Reference} & \multicolumn{3}{c|}{Clinical Evidence} & \multicolumn{2}{c|}{Clinical Case} & \multirow{2}{*}{\centering Avg.}\\
\cmidrule(r){1-1} \cmidrule(r){3-5} \cmidrule(r){6-8} \cmidrule(r){9-10}
Model& & Biology & Bioin. & MedS.  & MedE.  & MedD. & PMCT. & PMCC. & IIYiC.  \\
\midrule
\rowcolor{gray!8} \multicolumn{11}{l}{\textbf{Sparse Retrieval}} \\
\midrule
BM25~\cite{robertson2009probabilistic} &-  & 19.19 & 21.55& 19.68  & 0.66 & 2.55 & 23.69 & 21.66 & 12.02 & 15.13 \\
\midrule
\rowcolor{gray!8} \multicolumn{11}{l}{\textbf{Base Size (< 1B)}} \\
\midrule
Contriever~\cite{izacard2021unsupervised}&110M &9.15& 18.02  & 25.22  & 1.71 & 2.52 & 11.47 & 13.40 & 12.57 & 11.76\\
MedCPT\mtag~\cite{jin2023medcpt}&220M &2.15  & 17.57 & 14.74  & 1.68 & 2.02 & 11.33 & 14.62 & 8.03  & 9.02 \\
InstructOR-L~\cite{su2022one} &335M & 15.82 & 29.71 & 36.88 & 3.84 & 4.81 & 15.84 & 9.02  & 13.77 & 16.21\\
BGE-Large~\cite{xiao2024c} &335M & 12.71 & 27.04 & 27.76 & 4.10 & 8.33 & 26.45 & 15.06 & 14.72 & 17.02\\
BMRetriever\mtag~\cite{xu2024bmretriever} &410M & 12.37  & 29.92 & 31.26& 4.46 & 6.28 & 25.31 & 17.46 & 17.73 & 18.10\\
\midrule
\rowcolor{gray!8} \multicolumn{11}{l}{\textbf{Large Size (> 1B)}} \\
\midrule
InstructOR-XL~\cite{su2022one}&1.5B &21.56  & 32.91& 36.79 & 4.63 & 4.29 & 14.18 & 14.49 & 16.17 & 18.13\\
BMRetriever-2B\mtag~\cite{xu2024bmretriever} &2B & 19.50& 33.30 & 39.45  & 9.97 & 9.31 & 38.01 & 25.65 & 22.30 & 24.69\\
E5-mistral~\cite{wang2023improving} &7B & 18.81 & 42.86 & 41.77 & 6.70 & 11.54 & 23.58 & 31.17 & 22.93 & 24.92\\
BMRetriever-7B\mtag~\cite{xu2024bmretriever} &7B & 23.62  & 44.01& 44.91 & 11.55 & 16.95 &\underline{46.88} & 29.14 &\underline{24.36} & 30.18\\
SFR-Embedding~\cite{meng2024sfrembedding} &7B & 19.56  &\underline{45.91}& \underline{46.01} &\underline{11.98} &\underline{17.49} & 44.19 & 36.36 & 23.71 & 30.65\\
GritLM-7B~\cite{muennighoff2024generative} &7B & 24.99  & 43.98 &45.94&\textbf{12.32} &\textbf{19.86} & 39.88 &\underline{37.08} &\textbf{24.94} & \underline{31.12}\\
NV-Embed-v2~\cite{lee2024nv} &7B &\textbf{27.15}&\textbf{50.10} & \textbf{47.81}  & 10.90 & 16.72 & 44.05 &\textbf{39.91} & 14.81 & \textbf{31.43} \\
Voyage-3~\cite{voyageemb} &- &\underline{25.42}& 38.98 & 41.63  & 8.74 & 9.36 & 45.28 & 28.68 & 20.64 & 27.34\\
OpenAI-3-large~\cite{openaiemb} &- & 23.82  & 40.51 & 44.05& 11.78 & 15.01 &\textbf{47.43} & 28.87 & 17.12 & 28.57\\

\bottomrule
\end{tabular}
}
\caption{The performance of retrieval models on R2MED. We report nDCG@10 for eight datasets: Biology, Bioinformatics (Bioin.), Medical Sciences (MedS.), MedXpertQA-Exam (MedE.), MedQA-Diag (MedD.), PMC-Treatment (PMCT.), PMC-Clinical (PMCC.), IIYi-Clinical (IIYiC.). $\dag$ denotes medical retrievers. \textbf{Bold} and \underline{underline} indicate the best and second-best results on each dataset.}
\label{tab:main_results}

\end{table*}
\noindent \textbf{Existing retrieval systems perform poorly on R2MED.} 
As shown in Table~\ref{tab:main_results}, retrieval models across a wide range of sizes and architectures achieve uniformly low performance on R2MED, with the best-performing model (NV-Embed-v2) reaching only 31.43 nDCG@10. These retrievers are primarily trained on conventional semantic relevance datasets, rendering them ineffective for reasoning-intensive retrieval. Notably, BM25 performs on par with base-size dense retrievers, while large-size models (> 1B) consistently outperform smaller ones. Interestingly, medical retrievers such as BMRetriever-7B show no clear advantage over general-purpose retrievers like GritLM-7B or NV-Embed-v2, despite pretraining on large biomedical corpora. This may stem from differences in backbone architectures as well as the limitations of medical training corpora, which often lack reasoning-driven retrieval data. These results underscore the limitations of current retrieval systems in complex medical contexts and motivate the development of models better aligned with the demands of reasoning-driven retrieval.

\begin{figure}[htbp]
    \centering
    \includegraphics[width=0.8\linewidth]{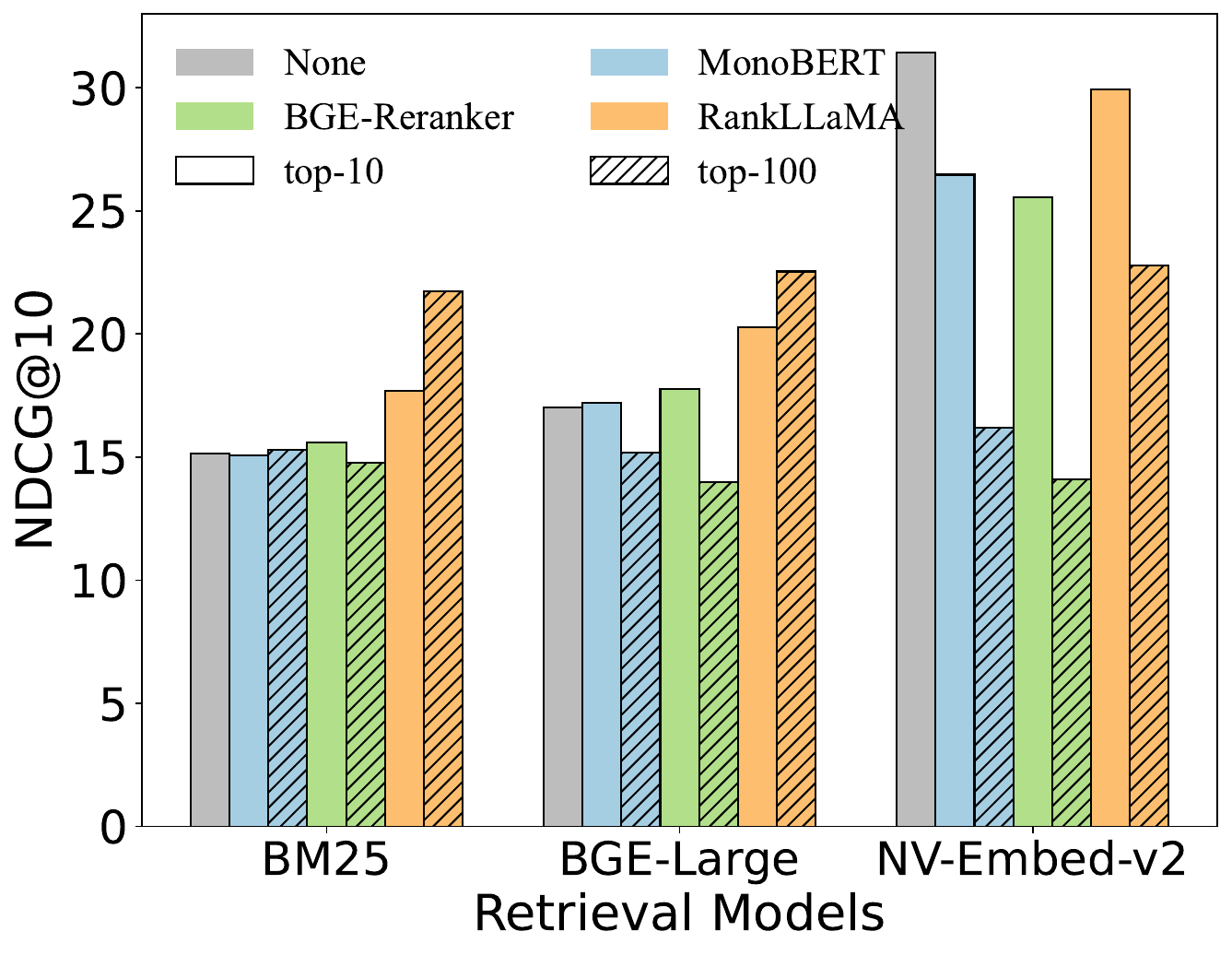}
    \caption{Average reranking performance on R2MED using three classic rerankers: MonoBERT, BGE-Reranker, and RankLLaMA. Detailed are in Table~\ref{tab:reranking_result}.}
    \label{fig:rerank-results}
\end{figure}

\noindent \textbf{Reranking methods offers inconsistent gains on R2MED.} 
Reranking has been a widely adopted strategy to improve retrieval performance \cite{nogueira2019multi,nogueira2020document,liu2025matryoshka}. We evaluate three representative rerankers, namely MonoBERT~\cite{nogueira2019multi}, BGE-Reranker-v2-m3~\cite{chen2024bge}, and RankLlama-7B~\cite{ma2024fine}, on the top-10 and top-100 documents retrieved by three retrievers. As shown in Figure~\ref{fig:rerank-results}, reranking yields clear improvements when the underlying retriever is relatively weak  (e.g., BM25 or BGE-Large), particularly in the top-10 setting. However, when applied to a stronger retriever like NV-Embed-v2, all three rerankers fail to deliver further gains and even degrade performance. Moreover, reranking over top-100 candidates proves substantially more difficult than over top-10, often leading to inconsistent or negative results. These findings indicate that reranking is not universally effective in reasoning-centric retrieval scenarios, and highlight the need for more robust reranking strategies tailored to the unique challenges posed by R2MED.


\begin{table}[htbp]
\centering

\small 
\setlength{\tabcolsep}{5pt} 
\begin{tabular}{lccc}
\toprule
\textbf{Methods} & BM25 & BGE. & NV. \\
\midrule
None & 15.13 & 17.02 & 31.43 \\
\midrule
HyDE$_{\text{Qwen-7B}}$   & 26.38 & 24.00 & 32.69 \\
HyDE$_{\text{Qwen-72B}}$  & 32.07 & 26.89 & 37.00 \\
HyDE$_{\text{GPT-4o}}$    & \textbf{37.70} & 28.63 & 39.37 \\
\midrule
Query2Doc$_{\text{Qwen-7B}}$  & 25.73 & 23.45 & 32.52 \\
Query2Doc$_{\text{Qwen-72B}}$ & 32.43 & 29.19 & 37.79 \\
Query2Doc$_{\text{GPT-4o}}$   & 35.85 & 30.72 & \textbf{41.66} \\
\midrule
LameR$_{\text{Qwen-7B}}$  & 25.32 & 25.44 & 34.52 \\
LameR$_{\text{Qwen-72B}}$ & 31.90 & 29.87 & 38.56 \\
LameR$_{\text{GPT-4o}}$   & 32.07 & \textbf{30.74} & 38.10 \\
\bottomrule
\end{tabular}
\caption{Average nDCG@10 score of generation-augmented retrieval (GAR) methods. \textbf{Bold} indicates the best results on each retriever. Detailed scores can be found in Table~\ref{tab:gar_result_bm25} to~\ref{tab:gar_result_nv}.}
\label{tab:gar_results}
\end{table}

\noindent \textbf{GAR methods demonstrate effectiveness on R2MED.} 
Recently, generation-augmented retrieval (GAR) methods, enhancing queries by leveraging LLMs to generate rewritten queries or hypothetical documents before retrieval, have emerged as a promising approach for adapting retrieval models to out-of-domain scenarios~\cite{mao2020generation, mao2024rafe, li2025reinforced}. We evaluate three representative GAR methods: HyDE~\cite{gao2022precise}, Query2Doc~\cite{wang2023query2doc}, and LameR~\cite{shen2023large}, each instantiated with three backbones of increasing capacity: Qwen2.5-7B-Instruct~\cite{qwen2.5}, Qwen2.5-72B-Instruct~\cite{qwen2.5}, and GPT-4o. As shown in Table~\ref{tab:gar_results}, larger generators consistently yield better retrieval performance, with GPT-4o achieving the highest scores across all three methods. Notably, Query2Doc with GPT-4o delivers the highest nDCG@10 of 41.66, significantly outperforming the best vanilla retriever. BM25 benefits most from GAR approachs, possibly due to its flexibility in handling out-of-distribution queries generated by LLMs. Overall, these results reinforce a central insight of R2MED that an intermediate answer serves as a crucial semantic bridge, effectively narrowing the gap between queries and relevant documents.

\section{Analysis}
\begin{table*}[t!]
\setlength{\tabcolsep}{3pt}

\centering
\resizebox{0.9\textwidth}{!}{
\small
\begin{tabular}{l|ccc|ccc|cc|c}
\toprule 
Model&  Biology & Bioin. & MedS.  & MedE.  & MedD. & PMCT. & PMCC. & IIYiC. &Avg.  \\
\midrule
NV-Embed-v2  &27.15 &50.1 &47.81 &10.90 &16.72 &44.05 &39.91 &14.81 &31.43 \\
\midrule
\multicolumn{10}{l}{\textbf{Large Language Model}} \\
\midrule
\rowcolor{green!10} Qwen2.5-7B-Ins. & 30.12 & 49.95 & 49.39 & 13.37 & 19.49 & 42.99 & 38.36 & 17.86 & 32.69\\
\rowcolor{gray!10} Qwen2.5-32B-Ins. &31.34 &52.35 &49.76 &16.40 &22.77 &45.31 &43.40 &21.35 &35.34\\
\rowcolor{blue!10} Llama3.1-70B-Ins. &31.21 &52.27 &51.19 &17.48 &27.53 &46.96 &46.90 &21.05 &36.82\\
GPT-4o &33.61 &54.15 &50.83 &23.08 &36.09 &47.35 &48.51 &21.30 &39.37\\
\midrule
\multicolumn{10}{l}{\textbf{Large Reasoning Model}} \\
\midrule
\rowcolor{gray!12} R1-Distill-Qwen-32B &33.1 &51.82 &49.39 &18.78 &27.38 &45.94 &42.16 &20.05 &36.08\\
\rowcolor{gray!15} QwQ-32B &32.26 &52.43 &49.91 &21.08 &31.29 &46.14 &41.06 &20.38 &36.82\\
\rowcolor{blue!12} R1-Distill-Llama-70B &32.83 &53.31 &50.32 &22.98 &33.78 &47.04 &46.53 &21.35 &38.52\\
\rowcolor{blue!17} HuatuoGPT-o1-70B &31.25 &52.81 &49.55 &25.25 &38.33 &48.93 &48.57 &21.77 &39.56\\
o3-mini &34.01& 55.9  & 51.28  & 28.99 & 40.30 & 48.97 &50.86 & 20.47 & 41.35\\
\midrule
\multicolumn{10}{l}{\textbf{Search-Enhanced Large Reasoning Model}} \\
\midrule
Search-R1-3B & 25.76	&47.53	&47.57	&11.98	&18.88	&45.66	&38.57	&17.95	&31.74\\
\rowcolor{green!15} Search-R1-7B & 30.84 & 50.66 & 49.07 & 15.05 & 20.46 & 47.36 & 45.49 & 21.95 & 35.11\\
\rowcolor{gray!20} Search-o1$_{\mathbf{QwQ\text{-}32B}}$ & 31.82 & 53.33 & 51.32 & 21.68 & 32.80 & 45.93 & 47.37 & 21.52 & 38.22\\
Search-o1$_{\mathbf{Qwen3\text{-}32B}}$ & 33.46 & 51.04 & 50.20 & 22.91 & 32.02 & 46.88 & 46.18 & 21.32 & 38.00\\
\bottomrule
\end{tabular}
}
\caption{LRM performance on R2MED. Rows of the same color indicate shared backbones (e.g., R1-Distill-Qwen-32B uses Qwen2.5-32B-Ins). See Tables \ref{tab:lrm_result_bm25}–\ref{tab:lrm_result_openai} for results with other retrievers.}
\label{tab:lrm_results}
\end{table*}
\subsection{LRMs Bring Marginal Gains on R2MED}
Recent advancements in large reasoning models (LRMs), such as OpenAI's o1~\cite{jaech2024openai} and DeepSeek-R1~\cite{guo2025deepseek}, have demonstrated strong performance on complex medical reasoning tasks~\cite{xie2024preliminary,jiang2025meds}. To assess their utility for reasoning-driven retrieval, we evaluate two paradigms: LRMs and search-enhanced LRMs. The LRM group includes DeepSeek-R1-Distill-Qwen-32B~\cite{guo2025deepseek}, QwQ-32B~\cite{qwq32b}, DeepSeek-R1-Distill-Llama-70B~\cite{guo2025deepseek}, HuatuoGPT-o1-70B~\cite{chen2024huatuogpt}, and o3-mini~\cite{o3-mini}. Search-enhanced LRMs incorporate agentic search workflows that enable dynamic retrieval of external knowledge during inference, particularly when the model encounters uncertainty. We evaluate Search-R1~\cite{jin2025search}, which is based on Qwen2.5-3b-it-em-ppo and Qwen2.5-7b-it-em-ppo, and Search-o1~\cite{li2025search}, implemented with QwQ-32B and Qwen3-32B~\cite{qwen3} as backbones. We use MedCorp~\cite{xiong2024benchmarking} as the retrieval corpus, with BM25 serving as the underlying search engine to ensure retrieval efficiency during reasoning. All models are evaluated under the HyDE setup, while only the final answer (excluding the reasoning trace) is extracted as the rewritten query. More details are provided in Appendix~\ref{app:evaluation_setting}. 

Experimental results in Table~\ref{tab:lrm_results} show that LRMs consistently outperform their base LLM counterparts across different backbones. For example, DeepSeek-R1-Distill-Llama-70B achieves an nDCG@10 of 38.52, surpassing Llama3.1-70B-Instruct’s 36.82. This trend holds across other model pairs, indicating that enhanced reasoning capabilities contribute modestly to improved retrieval performance on R2MED. Meanwhile, fine-tuned models on medical (e.g., HuatuoGPT-o1) also show slight gains. Notably, search-enhanced LRMs bring further gains by incorporating external knowledge during inference, for instance, Search-o1$\mathbf{_{QwQ-32B}}$ improves upon its base model QwQ-32B, raising nDCG@10 from 36.82 to 38.22. Search-o1 outperforms Search-R1 across multiple metrics, likely due to the incorporation of a reason-in-documents module that better utilizes retrieved content.

Despite these improvements, the overall gains remain modest, suggesting current LRMs have yet to fully realize their potential in reasoning-based retrieval. Additionally, these methods raise substantial efficiency concerns. LRMs generate long reasoning traces that increase token usage and latency, while search-enhanced models add computational overhead through multiple retrievals during generation. As such, it is crucial to assess LRMs through a balanced lens of both effectiveness and efficiency. Designing methods that jointly optimize for both remains an open and pressing challenge.

\subsection{Accurate Reasoning Leads to Better Retrieval}
\begin{figure}[htbp]
  \centering
  \includegraphics[width=0.8\linewidth]{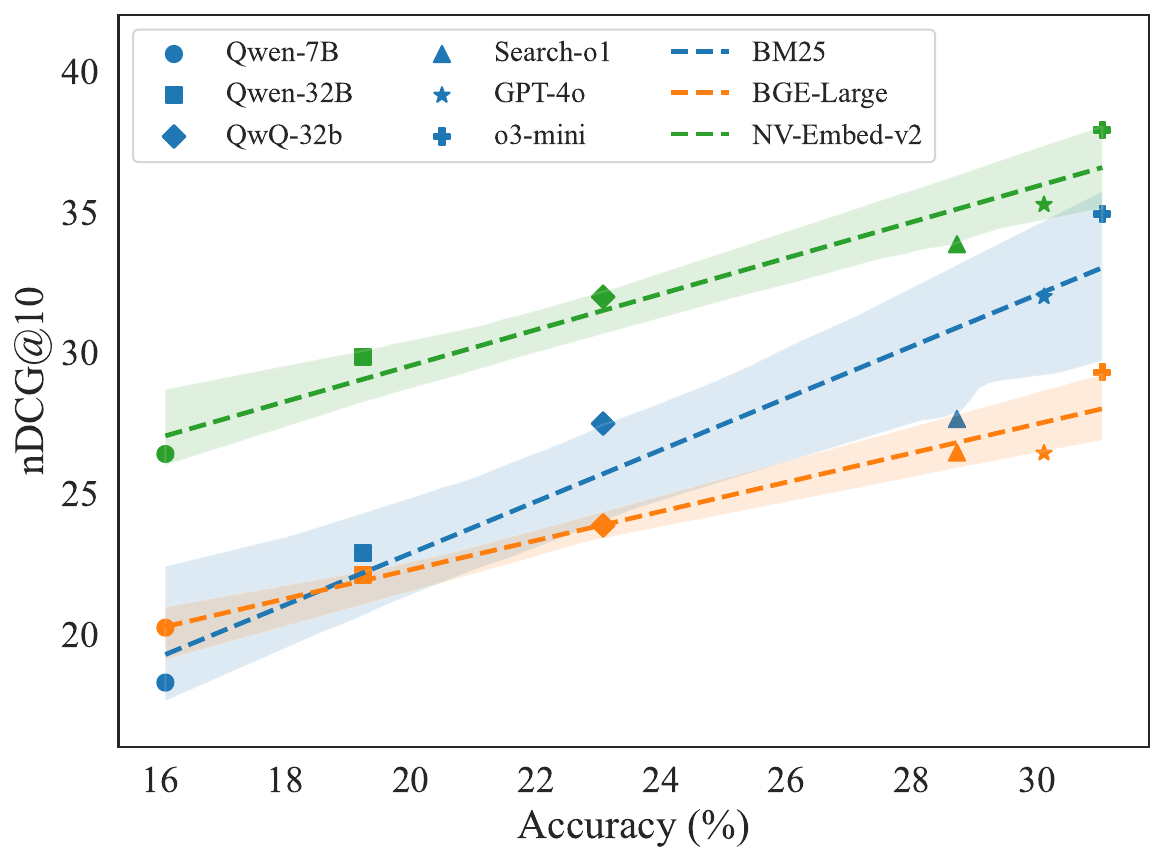}
  \caption{Correlation between reasoning answer accuracy and retrieval performance.}
  \label{fig:answer_accuracy}
\end{figure}
To gain deeper insights into how LRMs contribute to retrieval improvements on R2MED, we investigate the relationship between the accuracy of generated intermediate answers and the final retrieval performance. We focus on five datasets (excluding the Q\&A reference retrieval task), as these contain well-defined medical entities or concise phrases as golden answers, thereby allowing for more reliable evaluation. We evaluate six representative models: Qwen2.5-7B-Instruct, Qwen2.5-32B-Instruct, QwQ-32B, Search-o1$\mathbf{_{QwQ-32B}}$, GPT-4o, and o3-mini. For each model, we extract the predicted answer entity from its generated reasoning trace and assess its correctness using GPT-4o as a judge.

As shown in Figure~\ref{fig:answer_accuracy}, answer accuracy is strongly correlated with retrieval performance. Models that generate more accurate intermediate answers retrieve more relevant documents. Moreover, LRMs outperform size-matched LLMs in both answer accuracy and retrieval effectiveness, highlighting the benefits of long-chain reasoning. These results reinforce the reasoning-centric nature of R2MED, where higher answer accuracy directly contributes to better retrieval performance.

\section{Conclusion and Future Work}
We introduce R2MED, the first benchmark specifically designed for reasoning-driven retrieval in medicine. It comprises eight datasets spanning diverse clinical scenarios, including medical question answering and diagnostic reasoning. Our experiments reveal that existing retrievers perform poorly on R2MED, with the strongest model reaching only 31.4 nDCG@10. While large reasoning models can provide modest improvements (up to 41.4), a significant performance gap remains. R2MED reveals a fundamental challenge: effective retrieval in medicine requires reasoning, not just semantic matching. In the future, we plan to develop retrieval methods explicitly tailored for reasoning-driven retrieval tasks. Furthermore, we see promising opportunities in extending R2MED to multimodal medical retrieval, incorporating imaging data. Overall, we hope R2MED lays the groundwork for future research into retrieval systems that meet the complex reasoning demands of medical applications.

\newpage

\section{Limitations.}
While R2MED covers a diverse set of medical retrieval scenarios and is designed to enable reasoning-centric evaluation, it still has several limitations. First, the benchmark focuses exclusively on text-based retrieval and does not account for multimodal evidence such as medical images, laboratory results, or other structured clinical data, which are often essential in real-world medical decision-making. Second, query filtering, relevance annotation, and parts of the benchmark construction pipeline rely on large language models, whose outputs may be affected by limitations in medical knowledge, instruction-following ability, and model-specific biases. Although we adopt a two-stage expert review process to improve annotation quality and remove problematic instances, the resulting labels should not be regarded as entirely free of subjectivity or systematic error. Third, the relevant document mining stage depends on a limited set of retrievers and top-$k$ retrieval results, which means some truly relevant documents may be missed and remain unjudged due to model bias or recall limitations, potentially affecting evaluation outcomes. Finally, despite the diversity of sources and query types included in R2MED, the benchmark inevitably reflects source and distribution biases and cannot fully represent the full spectrum of medical information needs across institutions, populations, and specialties. More broadly, R2MED is intended to study reasoning-driven retrieval rather than to capture the complete workflow of clinical decision-making, which in practice involves longitudinal patient context, multimodal evidence, expert interpretation, and downstream action.

\section{Ethical Considerations}
To construct R2MED, we exclusively curated data from publicly available sources, including published medical resources, publicly accessible medical platforms, and anonymized clinician-authored case materials. No identifiable personal health information (PHI) was intentionally collected, used, or disclosed in the process, and we followed established data collection procedures to comply with applicable copyright and privacy requirements, including the removal or anonymization of personal identifiers where necessary. Nevertheless, because the benchmark contains medical content and clinical scenarios, it may still include medically sensitive or potentially distressing material. R2MED is developed and released solely for research and academic evaluation in a controlled setting, with the goal of studying reasoning-driven medical retrieval. It is not intended for clinical use, medical diagnosis, treatment planning, or real-world decision-making under any circumstances. Accordingly, outputs from models evaluated on R2MED must be interpreted with caution and should never be used as a substitute for professional medical advice or expert judgment.

\bibliography{custom}

\newpage

\appendix


\section{Dataset Construction}
\label{app:data_construction}
In this section, we provide more detailed information about the four stages of dataset construction. Tables~\ref{tab:data_license_query}-\ref{tab:data_license_doc} summarize the sources of queries and documents for each dataset. The number of queries across different construction stages can be found in Tables~\ref{tab:query_num_qa}-\ref{tab:query_num_case}.

\subsection{Data Collection}
\label{app:data_collection}
\paragraph{Q\&A Reference Retrieval Datasets}
For the Biology dataset, we directly adopt the version curated by BRIGHT~\cite{su2024bright}. To broaden domain coverage, we additionally construct two new datasets sourced from the Bioinformatics and Medical Sciences communities on StackExchange. We select posts where the accepted answer has received more than three upvotes and contains at least one external URL. For each selected post, we extract 1-2 linked webpages to serve as the initial positive documents.

To construct the initial negative pool, we use domain-relevant Wikipedia corpora distinct from the sources of positive documents to avoid content overlap. Specifically, we use the \textit{medicine\_wiki} corpus~\cite{burgerbee2025medicine} for the Medical Sciences dataset and the \textit{wiki\_medical\_terms}~\cite{gamino_wiki_medical_terms_2024} for the Bioinformatics dataset. All documents, from external webpages and Wikipedia, are segmented into smaller passages by sentence-level splitting and regrouped into chunks of approximately 128 tokens.

\paragraph{Clinical Evidence Retrieval Datasets}
The clinical evidence retrieval task comprises three datasets, each representing a critical stage in clinical decision-making: examination recommendation, disease diagnosis, and treatment planning. We construct these datasets based on three representative medical question-answering sources: MedXpertQA~\cite{zuo2025medxpertqa} (1,861 multiple-choice questions), MedQA~\cite{jin2021disease} (1,273 multiple-choice questions), and MedRBench\_Treat~\cite{qiu2025quantifying} (496 open-ended questions). To ensure that queries genuinely require medical reasoning, we apply a multi-stage filtering and reformulation pipeline:

\begin{itemize}[leftmargin=1.3em,labelsep=1.0em,itemindent=0em]
    \item \textbf{Task-based Filtering}. We use GPT-4o to annotate each question with a clinical task type (examination, diagnosis, or treatment) and discard those not belonging to the targeted categories. The instruction is shown in Figure~\ref{inst:tag_filter}.
    \item \textbf{Rule-based Filtering}. We further remove questions that do not require reasoning or whose answers are not specific medical entities. This step is automated via GPT-4o with rule-based guidance (see Figure~\ref{inst:rule_filter}).
    \item \textbf{Difficulty Filtering}. To retain only challenging questions, we evaluate each multiple-choice item using four small-scale instruction-tuned models: Qwen2.5-7B-Instruct, Llama3.1-8B-Instruct, Gemma-2-9B-it, and Qwen-14B-Instruct. Questions correctly answered by more than one model are filtered out.
    \item \textbf{Open-ended Reformulation}. Selected multiple-choice questions are reformulated into open-ended formats, with corresponding answers extracted using GPT-4o. The transformation prompt is shown in Figure~\ref{inst:rule_reformatting}.
\end{itemize}

For negative corpus construction, we use different sources for each dataset. For the examination dataset (MedXpertQA-Exam), we sample from the Wikipedia subset of the MedCorp corpus~\cite{xiong2024benchmarking}. For the diagnosis dataset (MedQA-Diag), we use medical textbook materials released with the original benchmark~\cite{jin2021disease}. For the treatment dataset (PMC-Treatment), we retain the original article associated with each question as the positive document. To build a challenging negative set, we crawl approximately 14,000 case reports from the PubMed Central Open Access (PMC OA) Subset~\footnote{\url{https://pmc.ncbi.nlm.nih.gov/tools/openftlist/}}, focusing on case reports tagged with diagnosis or treatment topics.

\begin{tcolorbox}[
    colframe = gray,       
    colback = gray!5!white,             
    coltitle = white,                   
    coltext = black,                    
    fonttitle = \bfseries,              
    title = Instruction for Task-based Filtering,  
    boxrule = 1pt,                      
    arc = 2mm,                          
    width = \linewidth,                 
    left = 7pt,                         
    right = 7pt,                        
    top = 3pt,                          
    bottom = 3pt                        
]
\fontsize{8.5pt}{10pt}\selectfont
You are a medical expert. Your task is to classify a given medical Question-Answer (QA) pair into one of the following categories based on the question's intent and the answer's entity type:
\\
\hspace*{1em}1. Examinations Recommendation (EXM): The question asks for the most appropriate diagnostic test or examination to confirm a suspected condition. The answer should be Laboratory Tests, Imaging Examinations, Endoscopic Examinations, or Other Examinations.\\
\hspace*{1em}2. Diagnostic Reasoning (DIA): The question asks for the most likely disease, syndrome, etiology, or functional disorder affecting the patient. The answer should be Disease Diagnosis, Syndrome Diagnosis, Etiological Diagnosis, or Functional Disorder Diagnosis.\\
\hspace*{1em}3. Treatment Planning (TRT): The question asks for the best treatment plan, including pharmacological, surgical, or preventive measures. The answer should be Pharmacological Treatment, Surgical Treatment, Other Therapies, and Preventive Measures.\\
\texttt{**Task:**}\\
For each given Question-Answer (QA) pair, determine the most appropriate classification from the three categories above. If the QA pair does not fit any category, return "Other".\\
\texttt{**Output Format:**}\\
Your output should follow the following format, do not output any additional content: \\
\hspace*{1em}- Classification: [one of EXM/DIA/TRT/Other] \\
\hspace*{1em}- Reasoning: [brief explanation of why this classification was chosen] \\
Here is the Question-Answer (QA) pair: \\
<Question>: \{QUESTION\} \\
<Answer>: \{ANSWER\} \\
\end{tcolorbox}
\captionof{figure}{Instruction for filtering questions based on the task.}
\label{inst:tag_filter}
\begin{tcolorbox}[
    colframe = gray,       
    colback = gray!5!white,             
    coltitle = white,                   
    coltext = black,                    
    fonttitle = \bfseries,              
    title = Instruction for Rule-based Filtering,  
    boxrule = 1pt,                      
    arc = 2mm,                          
    width = \linewidth,                 
    left = 7pt,                         
    right = 7pt,                        
    top = 5pt,                          
    bottom = 5pt                        
]
\fontsize{8.5pt}{10pt}\selectfont
You are an expert in filtering and evaluating multiple-choice questions for advanced reasoning tasks. Your task is to evaluate a given question and determine whether it meets the following criteria:\\
\hspace*{1em}1. Depth of Reasoning: The question should require deeper reasoning. If the question appears too simple, mark it as "Too Simple".\\
\hspace*{1em}2. Unambiguous Correct Answer: The question must have a unique and unambiguous correct answer. If the question asks for "incorrect options" or allows for multiple correct answers, mark it as "Ambiguous Answer".\\
\hspace*{1em}3. Open-Ended Reformulation Feasibility: The question should be suitable for reformatting into an open-ended format. If the question cannot be easily reformulated into an open-ended problem and a clear ground-truth answer, mark it as "Not Reformulatable".\\
\hspace*{1em}4. Medical Entity as the Correct Answer: The correct answer must be a medical entity, such as a disease, drug, symptom, anatomical structure, laboratory test, imaging examination, or treatment method. If the correct option is an abstract concept, behavior, tool, or any non-medical entity, mark it as "Non-Medical Entity". \\
For each question, provide one of the following evaluations:\\
 \hspace*{1em}- "Pass" (The question meets all the criteria.)\\
 \hspace*{1em}- "Too Simple"\\
 \hspace*{1em}- "Ambiguous Answer"\\
 \hspace*{1em}- "Not Reformulatable"\\
 \hspace*{1em}- "Non-Medical Entity"\\
\texttt{**Output Format:**}\\
Your output should follow the following format, do not output any additional content: \\
\hspace*{1em}- Evaluation: [one of the five evaluations]\\
\hspace*{1em}- Explanation: [Brief explanation]\\
Here is the Multiple-choice Question and its answer:\\
<Question>: \{QUESTION\} \{OPTION\} \\
<Answer>: \{ANSWER\} \\
\end{tcolorbox}
\captionof{figure}{Instruction for filtering questions based on rules.}
\label{inst:rule_filter}
\begin{tcolorbox}[
    colframe = gray,       
    colback = gray!5!white,             
    coltitle = white,                   
    coltext = black,                    
    fonttitle = \bfseries,              
    title = Instruction for Open-ended Reformulation,  
    boxrule = 1pt,                      
    arc = 2mm,                          
    width = \linewidth,                 
    left = 7pt,                         
    right = 7pt,                        
    top = 5pt,                          
    bottom = 5pt                        
]
\fontsize{8.5pt}{10pt}\selectfont
You will be given a multiple-choice clinical question. Your task is to transform it into an open-ended question while preserving the original language and structure as much as possible. \\
Follow these steps:\\
\hspace*{1em}1. Remove the multiple-choice options from the original question. \\
\hspace*{1em}2. If the original question contains phrases like “Which of the following...”, rewrite it into a self-contained open-ended form, but only minimally modify the wording required to make it complete without the options.\\
\texttt{**Output Format:**}\\
Your output should follow the following format, do not output any additional content: \\
\hspace*{1em}- Open-ended Question: [your rewritten question] \\
\hspace*{1em}- Standard Answer: [a concise standard answer] \\
Here is the Multiple-choice Question and its answer:\\
<Question>: \{QUESTION\} \{OPTION\} \\
<Answer>: \{ANSWER\} \\
\end{tcolorbox}
\captionof{figure}{Instruction for reformatting open-question.}
\label{inst:rule_reformatting}
\begin{tcolorbox}[
    colframe = gray,       
    colback = gray!5!white,             
    coltitle = white,                   
    coltext = black,                    
    fonttitle = \bfseries,              
    title = Instruction for Quality Filtering,  
    boxrule = 1pt,                      
    arc = 2mm,                          
    width = \linewidth,                 
    left = 7pt,                         
    right = 7pt,                        
    top = 5pt,                          
    bottom = 5pt                        
]
\fontsize{8.5pt}{10pt}\selectfont
You are given a medical article containing the following components: Title, Abstract, Case 1 (clinical case), and Case 2 (clinical case). Your task is to assess whether this article is suitable for constructing a similar-case retrieval dataset focused on disease diagnosis. \\
Evaluate the article using three independent criteria, each scored from 0 to 10, based on the following standards: \\
Scoring Scale (applies to all three criteria): \\
\hspace*{1em}10: Fully satisfies the criterion\\
\hspace*{1em}7–9: Mostly satisfies the criterion, only minor issues\\
\hspace*{1em}4–6: Partially satisfies the criterion, some key limitations\\
\hspace*{1em}1–3: Minimally satisfies the criterion, major gaps\\
\hspace*{1em}0: Does not satisfy the criterion at all\\
\texttt{**Scoring Criteria:**}\\
\hspace*{1em}Score 1 – Diagnostic Focus: Does the title and abstract indicate that the main goal of the article is to study or explain the diagnosis process or mechanism of a specific disease shared across the cases?\\
\hspace*{1em}Score 2 – Case Completeness \& Shared Diagnosis: Do both cases contain sufficient clinical details? Is the same diagnosis clearly stated in both cases?\\
\hspace*{1em}Score 3 – Diagnostic Similarity \& Supportiveness: Are the clinical presentations of the two cases similar? Does Case 2 provide diagnostic insight or evidence that could support the diagnosis in Case 1?\\
\texttt{**Output Format:**}\\
Your output should follow the following format, do not output any additional content: \\
\hspace*{1em}- Score 1: 0-10\\
\hspace*{1em}- Score 2: 0-10\\
\hspace*{1em}- Score 3: 0-10\\
\hspace*{1em}- Explanation: [An overall explanation]\\
Here is the article:\\
\{ARTICLE\}
\end{tcolorbox}
\captionof{figure}{Instruction for filtering cases based on quality.}
\label{inst:rule_quality}

\paragraph{Clinical Case Retrieval Datasets}
For the clinical case retrieval task, we collect patient case records from two primary sources: (1) PMC-Patients~\cite{zhao2023large}, a curated collection of multi-case clinical reports from PubMed Central, and (2) IIYi-Clinical, a dataset we construct by crawling 10k anonymized patient records from over ten departments on the IIYi online consultation platform. All collected records undergo strict de-identification and privacy-preserving processing. We use GPT-4o-mini to translate IIYi's data into the corresponding English version. Dataset construction follows a three-stage pipeline:

\begin{itemize}[leftmargin=1.3em,labelsep=1.0em,itemindent=0em]
    \item \textbf{Task Filtering}. For PMC-Patients, we identify multi-case articles and extract only the first described case in each as the query source. For IIYi-Clinical, we group patient records by diagnostic label using rule-based matching. One case is selected as the query, and the remaining cases within the same group constitute the candidate retrieval pool.
    \item \textbf{Quality Filtering}. Each candidate case is assessed by GPT-4o across three dimensions: diagnostic focus, case completeness, and diagnostic similarity. Only cases that meet predefined thresholds on all three criteria are retained (see Figure~\ref{inst:rule_quality}).
    \item \textbf{Question Formulation}. We use GPT-4o to extract the patient's clinical presentation from each full case, removing any diagnostic reasoning or outcome information(see Figure~\ref{inst:rule_rewriting}).
\end{itemize}

For each query case, we construct the initial positive set by selecting full case records that share the same diagnostic results. All remaining cases in the corpus are treated as initial negatives. In the PMC-Patients dataset, where articles often contain multiple related cases, we retain only 1–3 additional cases from the same report as positive and exclude the remaining ones in the report.
\begin{tcolorbox}[
    colframe = gray,       
    colback = gray!5!white,             
    coltitle = white,                   
    coltext = black,                    
    fonttitle = \bfseries,              
    title = Instruction for Question Formulation,  
    boxrule = 1pt,                      
    arc = 2mm,                          
    width = \linewidth,                 
    left = 7pt,                         
    right = 7pt,                        
    top = 5pt,                          
    bottom = 5pt                        
]
\fontsize{8.5pt}{10pt}\selectfont
You will be given a clinical case and its corresponding diagnosis (disease entities). Your task is to construct a <Query, Answer> pair in Chinese from the Case, focusing only on disease diagnosis.\\
Follow these steps:\\
\hspace*{1em}Step 1. Construct the Query from the Case: Rewrite the Case into a clinical question that simulates a physician seeking diagnostic guidance. Retain detailed clinical findings: symptoms, labs, imaging, history, etc. Remove any explicit diagnostic statements. Ensure the resulting description is clear, coherent, and self-contained. End the query with a natural disease diagnostic question.\\
\hspace*{1em}Step 2. Write the Answer: Extract the diagnostic reasoning and final diagnosis from the case. The answer should: Contain the reasoning steps or evidence used to reach the diagnosis. End with the confirmed disease entity as the final diagnosis. The disease names should match the given disease entity list.\\
\texttt{**Output Format:**}\\
Your output should follow the following format, do not output any additional content: \\
\hspace*{1em}- Query: Rewritten case 1 with diagnostic question \\
\hspace*{1em}- Answer: Diagnosis that answers the query \\
Here are two cases and their disease entities:\\
<Case>: \{CASE\}\\
<Disease>: \{DISEASE\}
\end{tcolorbox}
\captionof{figure}{Instruction for rewriting question.}
\label{inst:rule_rewriting}

\subsection{Relevant Document Mining}
\label{app:relevant_mining}
For each query, we use OpenAI o3 model to generate a step-by-step reasoning path, following the instructions detailed in Figure~\ref{inst:reasoning_path}. This yields a structured triplet \textless query, reasoning path, answer\textgreater, which we refer to as the multi-view retrieval set. To mine potentially relevant documents, we deploy a retrieval committee comprising BM25, MedCPT, and BGE-Large, ensuring complementary retrieval capabilities. Each element in the triplet is used independently as a retrieval query under each retriever, and the top-$100$ documents are retrieved.
\begin{tcolorbox}[
    colframe = gray,       
    colback = gray!5!white,             
    coltitle = white,                   
    coltext = black,                    
    fonttitle = \bfseries,              
    title = Instruction for Reasoning Path,  
    boxrule = 1pt,                      
    arc = 2mm,                          
    width = \linewidth,                 
    left = 7pt,                         
    right = 7pt,                        
    top = 5pt,                          
    bottom = 5pt                        
]
\fontsize{8.5pt}{10pt}\selectfont
Please analyze the given medical question and systematically reason through the problem to reach the correct answer.\\
<Question>:\\
\{QUESTION\}\\
<Correct Answer>: \{ANSWER\}\\
\end{tcolorbox}
\captionof{figure}{Instruction for generating reasoning path.}
\label{inst:reasoning_path}

\subsection{Relevance Assessment}
\label{app:relevance_assessment}
To evaluate the relevance between each query and its corresponding potential positive documents, we employ GPT-4o as the assessment model. The detailed instruction is illustrated in Figure~\ref{inst:relevance_assess_1}-\ref{inst:relevance_assess_3}. 
\begin{tcolorbox}[
    breakable,  
    enhanced,  
    colframe = gray,       
    colback = gray!5!white,             
    coltitle = white,                   
    coltext = black,                    
    fonttitle = \bfseries,              
    title = Instruction for Relevance Assessment (I),  
    boxrule = 1pt,                      
    arc = 2mm,                          
    width = \linewidth,                 
    left = 7pt,                         
    right = 7pt,                        
    top = 2pt,                          
    bottom = 5pt                        
]
\fontsize{8.5pt}{10pt}\selectfont
You are an expert in evaluating passages linked in medical Q\&A forum posts. Each post consists of a Question and a long Answer (containing reasoning and conclusions). 
You are given one such passage, and your task is to assess how well the passage supports or enhances understanding of the post.\\
Your evaluation consists of two independent scoring aspects, each rated on a 10-point scale:\\
Score 1. Coverage of Critical Information (10 points)\\
\hspace*{1em}10 points (Highly Relevant): Passage provides detailed and accurate information in the post.\\
\hspace*{1em}7-9 points (Moderately Relevant): Passage contains strong related information but may lack depth.\\
\hspace*{1em}4-6 points (Weakly Relevant): Passage vaguely touches on ideas in the post but in a general or indirect way.\\
\hspace*{1em}0-3 points (Not Relevant): Passage does not meaningfully overlap with the post content or is tangential.\\
Score 2. Contribution to the Answer’s Reasoning (10 points)\\
\hspace*{1em}10 points (Strong Support): The passage supports at least one critical reasoning step with detailed evidence.\\
\hspace*{1em}7-9 points (Moderate Support): The passage provides useful but incomplete support for a reasoning step.\\
\hspace*{1em}4-6 points (Weak Support): The passage only partially supports the reasoning process, with limited relevance.\\
\hspace*{1em}0-3 points (No Support): The passage does not contribute to the reasoning process in any meaningful way.\\
Then provide a clear explanation for your score, detailing which part(s) of the passage are connected to the answer, whether and how it contributes to understanding or reasoning.\\
\texttt{**Output Format:**}\\
Your output should follow the following format, do not output any additional content: \\
\hspace*{1em}- Score 1: [Score between 0 and 10 for the Relevance of Key Knowledge Points]\\
\hspace*{1em}- Score 2: [Score between 0 and 10 for the Support of the Reasoning Process]\\
\hspace*{1em}- Explanation: [Detailed explanation]\\
Here is the Post and Passage:\\
<Post>: \{QUESTION\}\\
<Passage>:\{PASSAGE\}\\
\end{tcolorbox}
\captionof{figure}{Instruction for relevance assessment on Q\&A reference retrieval datasets.}
\label{inst:relevance_assess_1}
\begin{tcolorbox}[
    colframe = gray,       
    colback = gray!5!white,             
    coltitle = white,                   
    coltext = black,                    
    fonttitle = \bfseries,              
    title = Instruction for Relevance Assessment (II),  
    boxrule = 1pt,                      
    arc = 2mm,                          
    width = \linewidth,                 
    left = 7pt,                         
    right = 7pt,                        
    top = 5pt,                          
    bottom = 5pt                        
]
\fontsize{8.5pt}{10pt}\selectfont
You are an expert in evaluating retrieved medical passages for their relevance and usefulness in supporting clinical reasoning. Your task is to assess how well a passage aligns with a given question-answer pair, its reasoning process, and the key knowledge points.\\ 
\texttt{**Scoring Criteria:**}\\
Score 1. Relevance to Key Knowledge Points (10 points):\\
\hspace*{1em}10 points (Highly Relevant): The passage provides detailed and precise information directly covering the key knowledge points.\\
\hspace*{1em}7-9 points (Moderately Relevant): The passage is strongly related to the key knowledge but lacks some depth.\\
\hspace*{1em}4-6 points (Weakly Relevant): The passage touches on the knowledge points but lacks detail.\\
\hspace*{1em}0-3 points (Not Relevant): The passage does not discuss the key knowledge points in a meaningful way or is only vaguely related.\\
Score 2. Support for the Reasoning Process (10 points):.\\
\hspace*{1em}10 points (Strong Support): The passage completely supports at least one critical reasoning step with detailed evidence.\\
\hspace*{1em}7-9 points (Moderate Support): The passage provides useful but incomplete support for a reasoning step.\\
\hspace*{1em}4-6 points (Weak Support): The passage only partially supports the reasoning process, with limited relevance.\\
\hspace*{1em}0-3 points (No Support): The passage does not contribute to the reasoning process in any meaningful way.\\
\texttt{**Output Format:**}\\
Your output should follow the following format, do not output any additional content: \\
\hspace*{1em}- Score 1: [Score between 0 and 10 for the Relevance of Key Knowledge Points]\\
\hspace*{1em}- Score 2: [Score between 0 and 10 for the Support of the Reasoning Process]\\
\hspace*{1em}- Explanation: [Detailed explanation]\\
Here is the Question-Answer (QA) pair, its reasoning process, the key knowledge points, and the medical passage:\\
<Question>: \{QUESTION\}, <Answer>: \{ANSWER\}, <Reason>: \{REASON\}\\
<Key\_point>:\{POINT\}\\
<Passage>:\{PASSAGE\}\\
\end{tcolorbox}
\captionof{figure}{Instruction for relevance assessment on clinical evidence retrieval datasets.}
\label{inst:relevance_assess_2}
\begin{tcolorbox}[
    colframe = gray,       
    colback = gray!5!white,             
    coltitle = white,                   
    coltext = black,                    
    fonttitle = \bfseries,              
    title = Instruction for Relevance Assessment (III),  
    boxrule = 1pt,                      
    arc = 2mm,                          
    width = \linewidth,                 
    left = 7pt,                         
    right = 7pt,                        
    top = 5pt,                          
    bottom = 5pt                        
]
\fontsize{8.5pt}{10pt}\selectfont
You are an expert in evaluating retrieved medical case reports for clinical diagnosis support. Your task is to assess how well a case aligns with a target clinical scenario and whether it supports the diagnostic reasoning process leading to the correct diagnosis.\\
\texttt{**Scoring Criteria:**}\\
Score 1. Diagnostic Match (10 points):\\
\hspace*{1em}10 points (Exact Match): The candidate case clearly states the same diagnosis as the answer. \\
\hspace*{1em}7-9 points (Close Match): The case suggests a clinically similar diagnosis or disease subtypes with the answer.\\
\hspace*{1em}4-6 points (Partially Related): The case may be related, but the diagnosis is unclear or partially related to the answer.\\
\hspace*{1em}0-3 points (Unrelated Diagnosis): The diagnosis is clearly different or unrelated.\\
Score 2. Diagnostic Reasoning Support (10 points):\\
\hspace*{1em}10 points (Strong Support): The case contains highly similar symptoms, findings, or patterns that mirror key parts of the reasoning process. \\
\hspace*{1em}7-9 points (Moderate Support): The case overlaps with some reasoning elements but lacks full coverage.  \\
\hspace*{1em}4-6 points (Weak Support): The case only partially relates to the reasoning, with limited diagnostic value.\\
\hspace*{1em}0-3 points (No Support): The case does not help with diagnostic reasoning in a meaningful way.\\
\texttt{**Output Format:**}\\
Your output should follow the following format, do not output any additional content: \\
\hspace*{1em}- Score 1: [Score between 0 and 10 for Diagnostic Match]\\
\hspace*{1em}- Score 2: [Score between 0 and 10 for Diagnostic Reasoning Support]\\
\hspace*{1em}- Explanation: [Detailed explanation]\\
Here is the Question-Answer (QA) pair, its reasoning process, the candidate case:\\
<Question>: \{QUESTION\}, <Answer>: \{ANSWER\}, <Reason>: \{REASON\}\\
<Case>:\{CASE\}\\
\end{tcolorbox}
\captionof{figure}{Instruction for relevance assessment on clinical case retrieval datasets.}
\label{inst:relevance_assess_3}

\subsection{Expert Review}
\label{app:expert_review}
To further ensure the clinical validity and quality of our benchmark, we conduct a two-stage expert review of all examples. 
In the first stage, a medically trained annotator (a PhD student) reviews the entire dataset. The annotator receives targeted training prior to annotation, including task-specific guidelines, calibration on example cases, and discussions with clinical experts. In the second stage, a medical expert reviews only the examples flagged as problematic and provides final judgments. 
Before annotation, we explained the usage of data and obtained consent from annotators.
Each example is evaluated based on three criteria:
\begin{itemize}[leftmargin=1.3em,labelsep=1.0em,itemindent=0em]
    \item \textbf{Completeness and Coherence of Reformulated Queries.} This criterion assesses whether the reformulated query (if applicable) is self-contained, clinically coherent, and provides sufficient clinical detail for a clinician to answer it. Incomplete or incoherent queries may lack critical patient information or pose ill-formed clinical questions.
    \item \textbf{Plausibility of the Reasoning Path.} This dimension evaluates whether the model-generated reasoning path reflects medically sound logic. High-quality reasoning paths adhere to accepted diagnostic or therapeutic pathways, maintain clinical plausibility, and avoid unsupported or medically invalid inferences.
    \item \textbf{Supportiveness of Positive Documents.} This assesses whether the positive documents provide sufficient and relevant evidence to support the query-answer pair. Strong supporting documents either directly present or clearly imply the necessary findings, differentials, or treatment considerations. Documents lacking topical relevance or clinical substance receive lower ratings.
\end{itemize}

\begin{figure*}[t]
\includegraphics[width=\linewidth]{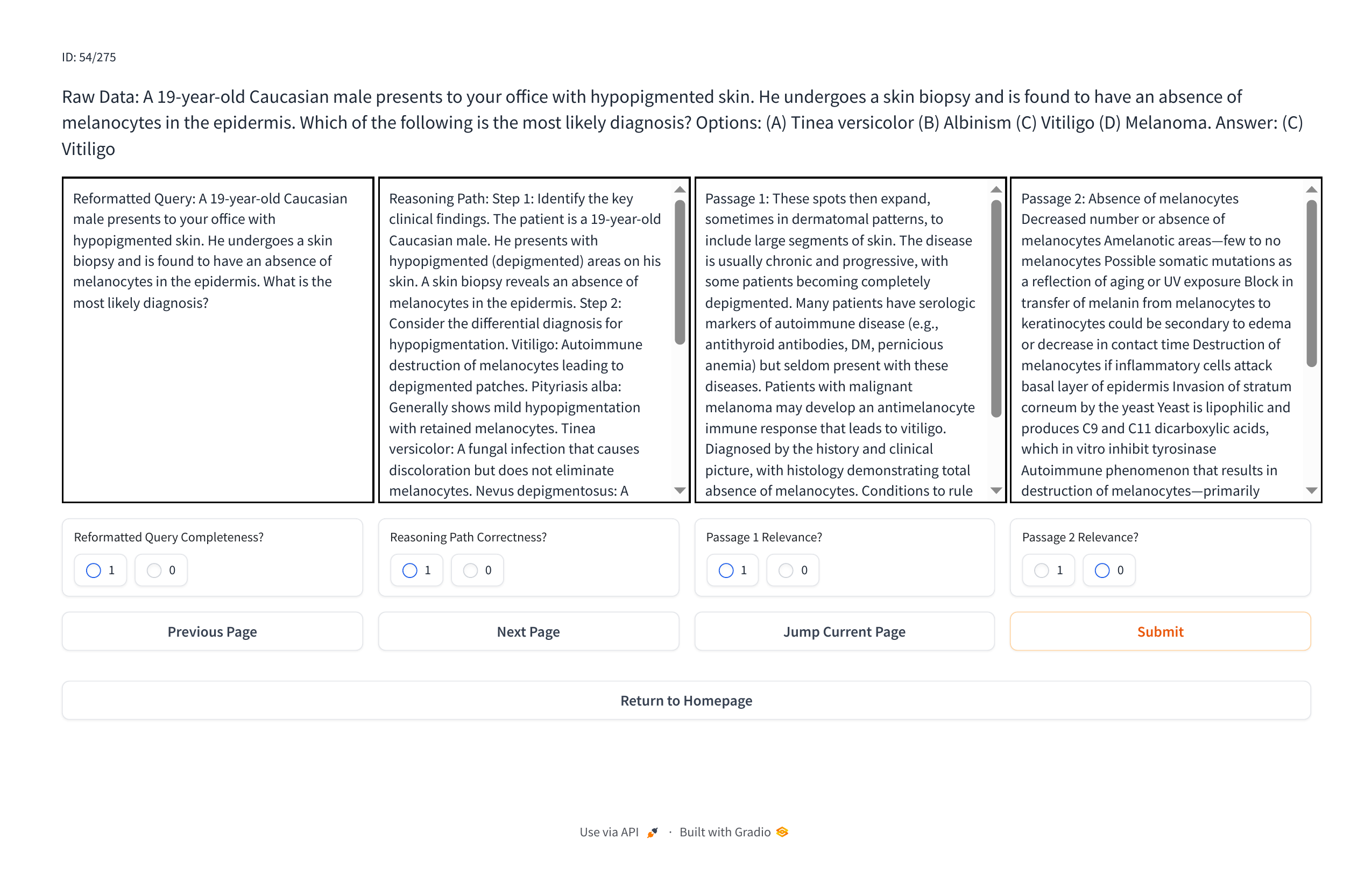}
  \caption{Annotation interface of R2MED.}
  \label{fig:data_anno}
\end{figure*}

In total, we review 833 query–answer pairs and approximately 2,500 associated positive documents across the seven datasets. Figure~\ref{fig:data_anno} illustrates an example from our annotation platform. The evolution of query counts through different processing stages is summarized in Table~\ref{tab:query_num_qa}-\ref{tab:query_num_case}. Fewer than 10\% of examples in each dataset are excluded after expert review. Common reasons for exclusion include: (1) exam-style or closed-ended queries such as “which of the following is...”; (2) supporting documents that are only loosely related to the query and lack substantive detail and reasoning paths that contain hallucinations; and (3) reasoning paths that contain factual hallucinations. Importantly, in the last case, if the associated positive documents remain clinically relevant, the example is retained despite imperfections in the reasoning path, as the documents still provide value for evaluating retrieval performance.

\begin{table}[t!]
\setlength{\tabcolsep}{3pt}
\caption{Number of queries during different stages on Q\&A reference task.}
\centering
\resizebox{\linewidth}{!}{
\begin{tabular}{l|ccc}
\toprule 
Dataset &Init. Query & AI Assessment &Expert Review  \\
\midrule
Biology &-  &- & 103 \\
Bioinformatics &104  &83 &77 \\
Medical Sciences &105  &95 & 88 \\
\bottomrule
\end{tabular}
}
\label{tab:query_num_qa}
\end{table}
\begin{table*}[t!]
\setlength{\tabcolsep}{3pt}
\caption{Number of queries during different stages on clinical evidence retrieval task.}
\centering
\resizebox{\textwidth}{!}{
\begin{tabular}{l|ccccccc}
\toprule 
Dataset &Init. Query &Task Filter & Rule Filter & Difficulty Filter & Reformatting & AI Assessment &Expert Review  \\
\midrule
MedXpertQA-Exam &1861	&302	&220	&181	&125 &105	&97 \\
MedQA-Diag &1273	&656	&543	&149	&126 &124	&118 \\
PMC-Treatment &496	&496	&496	&243	&243	&162 &150 \\
\bottomrule
\end{tabular}
}
\label{tab:query_num_evidence}
\end{table*}
\begin{table*}[t!]
\setlength{\tabcolsep}{3pt}
\caption{Number of queries during different stages on clinical case retrieval task.}
\centering
\resizebox{\textwidth}{!}{
\begin{tabular}{l|cccccc}
\toprule 
Dataset &Init. Query &Task Filter & Quality Filter & Rewriting & AI Assessment &Expert Review  \\
\midrule
PMC-Clinical &167k	&2817	&158	&158	&123	&114\\
IIYi-Clinical &10k	&553	&148	&148	&141	&129 \\
\bottomrule
\end{tabular}
}
\label{tab:query_num_case}
\end{table*}

\section{Data Examples}
\label{app:data_example}
In Table~\ref{tab:example_biology}-\ref{tab:example_iiyi}, we show more examples in R2MED.

\section{Data Diversity Analysis}
\label{app:data_diversity_analysis}
\begin{figure*}[t]
\includegraphics[width=\linewidth]{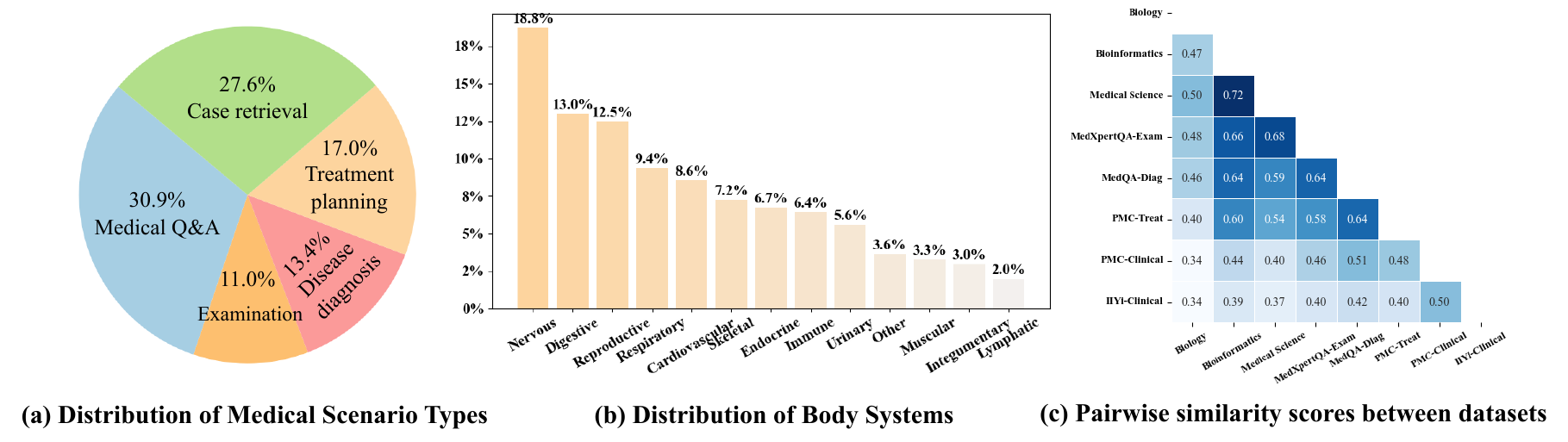}
  \caption{Attribute distributions of R2MED showcase its diversity and comprehensiveness.}
  \label{fig:bench_att}
\end{figure*}

We use GPT-4o to assign each query to one of twelve pre-defined body systems, based on the prompt shown in Figure~\ref{inst:system_annotation}.  Since the three StackExchange-derived datasets focus on general biomedical topics rather than clinical case scenarios, we exclude them from this annotation process and apply the labeling only to the remaining five datasets. In addition, we follow BEIR~\cite{thakur2021beir} and compute pairwise weighted Jaccard similarity scores between datasets to evaluate corpus-level distributional diversity. Each corpus is tokenized using the GPT-2 tokenizer, and overlap is measured at the token level. Low inter-dataset similarity confirms that R2MED spans heterogeneous distributions, presenting a strong generalization challenge for retrieval models.
The attribute distributions of R2MED is shown in Figure~\ref{fig:bench_att}.

\begin{tcolorbox}[
    colframe = gray,       
    colback = gray!5!white,             
    coltitle = white,                   
    coltext = black,                    
    fonttitle = \bfseries,              
    title = Instruction for Body System Annotation,  
    boxrule = 1pt,                      
    arc = 2mm,                          
    width = \linewidth,                 
    left = 7pt,                         
    right = 7pt,                        
    top = 5pt,                          
    bottom = 5pt                        
]
\fontsize{8.5pt}{10pt}\selectfont
You are an experienced medical doctor and independent practitioner. Your task will be to label a medical question according to the human body system it corresponds to.\\
You will be given a list of human body systems, followed by a medical question. Please determine which system the question best pertains to. If the question is related to multiple systems, only select the most relevant one.\\
Directly output the name of the final system you selected from the list of available systems, do not output any additional content.\\
\texttt{**Systems:**}\\
Integumentary, Skeletal, Muscular, Nervous, Endocrine, Cardiovascular, Lymphatic, Respiratory, Digestive, Urinary, Reproductive, Immune, Other\\
\texttt{**Question:**}\\
\{QUESTION\}\\
\end{tcolorbox}
\captionof{figure}{Instruction for body system annotation.}
\label{inst:system_annotation}

\section{Dataset License and Usage}
\subsection{Dataset License}
Table~\ref{tab:data_license_query}-\ref{tab:data_license_doc} summarize the data sources and corresponding licenses for the eight datasets included in R2MED. Most datasets are distributed under permissive licenses, such as variants of the Creative Commons Attribution (CC-BY) license and the MIT license, allowing sharing and adaptation for academic and research purposes. Although the MedRBench (PMC-Treatment) dataset does not explicitly specify a license in its repository, it is derived from PubMed Central Open Access (PMC OA) Subset~\footnote{\url{https://pmc.ncbi.nlm.nih.gov/tools/openftlist/}}, which is a publicly available resource widely used in academic research. For IIYi-Clinical, the data originates from a publicly accessible medical consultation platform. Prior
studies~\cite{yim2024overview,fan2024ai,hou2024msdiagnosis,jia2024medikal} confirm that data from this platform, once anonymized, is permissible for research and educational use. In summary, all datasets used in R2MED have been verified to be legally suitable for research. 

\subsection{Dataset Instance Metadata}
The codes of R2MED dataset is organized into three files: 
\texttt{query.jsonl}, \texttt{corpus.jsonl}, and \texttt{qrels.jsonl}, corresponding to queries, corpus passages, and relevance labels, respectively. Each file follows a line-delimited JSON (\texttt{.jsonl}) format. The schema for each file is summarized below:

\noindent\textbf{\texttt{query.jsonl}} 
Each row represents a query:
\begin{itemize}[leftmargin=3em,labelsep=1.0em,itemindent=0em]
    \item \texttt{id}: The unique identifier of the query.
    \item \texttt{text}: The textual content of the query.
    \item \texttt{answer}: The intermediate reasoning answer associated with the query.
    \item \texttt{doc\_id}: A list of golden positive document IDs that are relevant to the query.
    \item \texttt{body\_system}: The body system category associated with the query.
\end{itemize}

\noindent\textbf{\texttt{corpus.jsonl}}
Each row represents a document:
\begin{itemize}[leftmargin=3em,labelsep=1.0em,itemindent=0em]
    \item \texttt{id}: The unique identifier of the document.
    \item \texttt{text}: The textual content of the document.
\end{itemize}

\noindent\textbf{\texttt{qrels.jsonl}} 
Each row represents a query-passage relevance label:
\begin{itemize}[leftmargin=3em,labelsep=1.0em,itemindent=0em]
    \item \texttt{q\_id}: The ID of the query.
    \item \texttt{p\_id}: The ID of the document.
    \item \texttt{score}: The binary relevance score.
\end{itemize}

\subsection{Author Statement}
We affirm that all datasets incorporated into R2MED have been verified to originate from sources with open-source or permissive licenses (e.g., CC-BY, MIT). Nonetheless, we fully acknowledge the importance of respecting the rights and concerns of original data providers. Should any licensing issues be identified or brought to our attention, we are committed to responding promptly and taking appropriate corrective actions. To ensure transparency and traceability, we will maintain versioned releases of R2MED on both HuggingFace and GitHub, and ensure that any future updates (e.g., correction of metadata, expansion of sources).



\begin{table}[t]
\setlength{\tabcolsep}{3pt}
\caption{The license of query sources in R2MED.}
\centering
\resizebox{\linewidth}{!}{
\begin{tabular}{l|c}
\toprule 
Dataset & Source License \\
\midrule
Biology & CC-BY 4.0 license \\
Bioinformatics & CC-BY-SA license \\
Medical Sciences & CC-BY-SA license \\
MedXpertQA-Exam & MIT license \\
MedQA-Diag & MIT license \\
PMC-Treatment & - \\
PMC-Clinical & CC-BY-NC-SA 4.0 license \\
IIYi-Clinical & - \\
\bottomrule
\end{tabular}
}
\label{tab:data_license_query}
\end{table}


\begin{table}[t!]
\setlength{\tabcolsep}{3pt}
\caption{The license of document sources in R2MED.}
\centering
\resizebox{\linewidth}{!}{
\begin{tabular}{l|c}
\toprule 
Dataset & Source License \\
\midrule
Biology & CC-BY 4.0 license \\
Bioinformatics & GPL 3.0 license \\
Medical Sciences & CC-BY license \\
MedXpertQA-Exam & Public Domain (US Government Work) \\
MedQA-Diag & MIT license \\
PMC-Treatment & CC BY license \\
PMC-Clinical & CC-BY-NC-SA 4.0 license \\
IIYi-Clinical & - \\
\bottomrule
\end{tabular}
}
\label{tab:data_license_doc}
\end{table}

\section{Experiment Details}
\label{app:experiment_details}
\subsection{Model Details}
\label{app:model_details}
We summarize all retrieval and reranking models used in this study in Table~\ref{tab:retrieval_model_deatails}, including model names, parameter sizes, and implementation sources. The BM25 baseline is implemented using Pyserini~\cite{lin2021pyserini}. For details regarding the large language models and large reasoning models evaluated throughout the paper, please refer to Table~\ref{tab:other_model_deatails}.

In this work, we evaluate three generation-augmented retrieval (GAR) methods. HyDE~\cite{gao2022precise} prompts an instruction-following LLM in a zero-shot setting to generate a hypothetical answer document, which is then used to retrieve relevant information. Query2doc~\cite{wang2023query2doc} adopts a few-shot prompting strategy to generate pseudo-documents from the query using an LLM and expands the query with these documents to improve retrieval performance. LameR~\cite{shen2023large} augments queries by incorporating potential in-domain answers and prompting an LLM to rewrite the query in a retrieval-friendly form. For search-enhanced large reasoning models, we explore two recent approaches. Search-R1~\cite{jin2025search} extends DeepSeek-R1 by employing reinforcement learning to enable the model to autonomously generate multiple search queries and retrieve external evidence during multi-step reasoning. In contrast, Search-o1~\cite{li2025search} introduces an agent-based retrieval-augmented reasoning framework, incorporating a reason-in-documents module that iteratively refines the evidence selection throughout the reasoning process.

\begin{table}[t!]
\setlength{\tabcolsep}{3pt}
\caption{Detailed information on all of the retrieval and reranking models in our paper.}
\centering
\resizebox{\linewidth}{!}{
\begin{tabular}{l|c|c}
\toprule 
Model & Size & Architecture \\
\midrule
\rowcolor{gray!8}\multicolumn{3}{l}{\textbf{Retrieval Models}} \\
\midrule
BM25~\cite{robertson2009probabilistic} & N/A & Sparse \\
Contriever~\cite{izacard2021unsupervised} & 110M & Encoder \\
MedCPT~\cite{jin2023medcpt} & 220M & Encoder \\
InstructOR-L~\cite{su2022one} & 335M & Encoder \\
BGE-Large~\cite{xiao2024c} & 335M & Encoder \\
BMRetriever~\cite{xu2024bmretriever} & 410M & Encoder \\
InstructOR-XL~\cite{su2022one} & 1.5B & Encoder \\
BMRetriever-2B~\cite{xu2024bmretriever} & 2B & Decoder \\
E5-mistral~\cite{wang2023improving} & 7B & Decoder \\
BMRetriever-7B~\cite{xu2024bmretriever} & 7B & Decoder \\
SFR-Embedding~\cite{meng2024sfrembedding} & 7B & Decoder \\
GritLM-7B~\cite{muennighoff2024generative} & 7B & Decoder \\
NV-Embed-v2~\cite{lee2024nv} & 7B & Decoder \\
Voyage-3~\cite{voyageemb} & N/A & Dense \\
OpenAI-3-large~\cite{openaiemb} & N/A & Dense \\
\midrule
\rowcolor{gray!8}\multicolumn{3}{l}{\textbf{Reranking Models}} \\
\midrule
MonoBERT~\cite{nogueira2019multi} & 335M & Encoder \\
BGE-Reranker~\cite{chen2024bge} & 568M & Encoder \\
RankLLaMA~\cite{ma2024fine} & 7B & Decoder \\
\bottomrule
\end{tabular}
}
\label{tab:retrieval_model_deatails}
\end{table}

\begin{table}[t!]
\setlength{\tabcolsep}{3pt}
\caption{All LLMs and LRMs used in experiments.}
\centering
\resizebox{\linewidth}{!}{
\begin{tabular}{l|c}
\toprule 
Model & Size \\
\midrule
\rowcolor{gray!8}\multicolumn{2}{l}{\textbf{Large Language Models}} \\
\midrule
Qwen2.5-7B-Ins.~\cite{qwen2.5} & 7B \\
Qwen2.5-32B-Ins.~\cite{qwen2.5} & 32B \\
Qwen2.5-72B-Ins.~\cite{qwen2.5} & 72B \\
Llama3.1-70B-Ins.~\cite{grattafiori2024llama} & 70B \\
GPT-4o~\cite{achiam2023gpt} & N/A \\
\midrule
\rowcolor{gray!8}\multicolumn{2}{l}{\textbf{Large Reasoning Models}} \\
\midrule
R1-Distill-Qwen-32B~\cite{guo2025deepseek} & 32B \\
QwQ-32B~\cite{qwq32b} & 32B \\
Qwen3-32B~\cite{qwen3} & 32B \\
R1-Distill-Llama-70B~\cite{guo2025deepseek} & 70B \\
HuatuoGPT-o1-70B~\cite{chen2024huatuogpt} & 70B \\
o3-mini~\cite{o3-mini} & N/A \\
\bottomrule
\end{tabular}
}
\label{tab:other_model_deatails}
\end{table}

\subsection{Evaluation Settings and Instructions}
\label{app:evaluation_setting}
We outline the evaluation instructions used for InstructOR-L, InstructOR-XL, BGE-Large, BMRetriever-410M/2B/7B, E5-mistral, SFR-Embedding, NV-Embed-v2, and GritLM-7B in Table~\ref{tab:retriever_instruction}. For the embedding model provided by Voyage, we specify the  "input\_type" parameter as either "query" or "document" to distinguish queries from documents.

In our experiments, the large language model generates one hypothetical document per query for HyDE. For Query2doc, we manually select two additional in-domain examples from each dataset to construct few-shot prompts as contextual guidance. LameR first retrieves the top-$10$ documents using BM25. These retrieved documents are then incorporated into the prompt as context to enhance pseudo-document generation quality. The exact prompts for all three GAR methods are provided in Table~\ref{tab:gar_instruction}.

For search-enhanced reasoning methods, we follow the official evaluation settings of Search-R1~\cite{jin2025search} and Search-o1~\cite{li2025search}. Both systems operate on MedCorp~\cite{xiong2024benchmarking}, a large-scale medical corpus containing 53.9 million document chunks. We set the maximum number of retrieval rounds to 10. In each round, the system retrieves the top-$5$ documents, which are passed into the reasoning model to iteratively synthesize evidence and generate final answers. All reported results are based on a single run. For embedding-based retrievers, the retrieval process is deterministic given fixed queries, documents, and model checkpoints, and thus does not involve meaningful run-to-run randomness. For LLM-based generation components (e.g., HyDE, Query2doc, and reasoning-based methods), we use low-temperature decoding to minimize stochastic variation and report the results from one run for each setting.

\subsection{Computing Resources}
\label{app:computing_resources}
All experiments were conducted on a machine with 4 NVIDIA A100 GPUs (40GB each). BM25 was evaluated on CPU, while all other retrieval models utilized GPU resources. Evaluation time varied according to model scale and complexity. For the retrievers presented in this paper, end-to-end evaluation of a single model generally requires no more than 8 hours using the 4 A100 GPUs. For methods involving large language model generation, such as pseudo-document generation or reasoning, we leverage vLLM~\cite{kwon2023efficient} to accelerate the inference process and reduce latency.

\section{More Experiment Results}
\begin{table*}[t!]
\centering
\caption{Instructions used for benchmarking different datasets for retrieval models.}
\resizebox{\textwidth}{!}{
\begin{tabular}{l|l}
\toprule
\rowcolor{gray!8}\multicolumn{2}{l}{\textbf{Instruction Templates for Different Retrievers}} \\
\midrule
Retriever & Instruction\\
\midrule
\multirow{2}{7em}{InstructOR-L, InstructOR-XL} & \textbf{Query:} Represent the \{$\mathrm{tag_1}$\} for retrieving relevant paragraphs:\\
 & \textbf{Doc:} Represent the \{$\mathrm{tag_1}$\} paragraph for retrieval:\\
 \midrule
 BGE-Large & \textbf{Query:} Represent this sentence for searching relevant passages:\\
 \midrule
 \multirow{2}{7em}{BMRetriever-410M/2B/7B} & {\textbf{Query:} Given a \{$\mathrm{tag_1}$\}, retrieve relevant passages that help answer the \{$\mathrm{tag_2}$\}\textbackslash$ \mathrm{n}$ Query:} \\ 
 & \textbf{Doc:} Represent this passage\textbackslash $\mathrm{n}$ passage: \\
 \midrule
\multirow{3}{7em}{E5-mistral, SFR-Embedding, NV-Embed-v2} & \multirow{3}{35em}{\textbf{Query:} Instruct: Given a \{$\mathrm{tag_1}$\}, retrieve relevant passages that help answer the \{$\mathrm{tag_2}$\}\textbackslash$ \mathrm{n}$ Query: }\\ \\ \\
 \midrule
\multirow{3}{7em}{GritLM-7B} & {\textbf{Query:} $<|\mathrm{user}|>$\textbackslash$ \mathrm{n}$ Given a \{$\mathrm{tag_1}$\}, retrieve relevant passages that help} \\ 
& answer the \{$\mathrm{tag_1}$\}$<|\mathrm{embed}|>$\textbackslash$ \mathrm{n}$ \\
 & \textbf{Doc:} $<|\mathrm{embed}|>$\textbackslash$ \mathrm{n}$\\
 \midrule
 \rowcolor{gray!8}\multicolumn{2}{l}{\textbf{Tags for Different Datasets}} \\
\midrule
Dataset & Tag\\
\midrule
Biology & $\mathrm{tag_1}$: Biology Post, $\mathrm{tag_2}$: post\\
\midrule
Bioinformatics & $\mathrm{tag_1}$: Bioinformatics Post, $\mathrm{tag_2}$: post\\
\midrule
Medical Sciences & $\mathrm{tag_1}$: Medical Sciences Post, $\mathrm{tag_2}$: post\\
\midrule
MedXpertQA-Exam & $\mathrm{tag_1}$: Medical Exam, $\mathrm{tag_2}$: exam\\
\midrule
MedQA-Diag & $\mathrm{tag_1}$: Medical Exam, $\mathrm{tag_2}$: exam\\
\midrule
PMC-Treatment & $\mathrm{tag_1}$: Clinical Case, $\mathrm{tag_2}$: case\\
\midrule
PMC-Clinical & $\mathrm{tag_1}$: Clinical Case, $\mathrm{tag_2}$: case\\
\midrule
IIYi-Clinical & $\mathrm{tag_1}$: Clinical Case, $\mathrm{tag_2}$: case\\
\bottomrule
\end{tabular}
}
\label{tab:retriever_instruction}
\end{table*}
\begin{table*}[t!]
\centering
\caption{Instructions used for evaluating different datasets for generation-augmented retrieval (GAR) methods. \{TEXT\}, \{EXAMPLE\}, and \{PASSAGE\} are the corresponding placeholder.}
\resizebox{\textwidth}{!}{
\begin{tabular}{l|l}
\toprule
Method & Instruction\\
\midrule
\rowcolor{gray!8}\multicolumn{2}{l}{\textbf{Biology, Bioinformatics, and Medical Sciences datasets}} \\
\midrule
HyDE & Please write a passage to help answer the \{domain\} post.$\textbackslash \mathrm{n}$ Post: \{TEXT\}$\textbackslash \mathrm{n}$ Passage: \\
\midrule
\multirow{2}{5em}{Query2Doc} &Write a passage that help answer the \{domain\} post.$\textbackslash \mathrm{n}$ Examples: \{EXAMPLE\} \\
&Real Test:$\textbackslash \mathrm{n}$ Post: \{TEXT\}$\textbackslash \mathrm{n}$ Passage: \\
\midrule
\multirow{2}{5em}{LameR} &Give a \{domain\} post and its possible relevant passages (most of these passages are wrong). Please write a\\
&correct passage that help answer the post.$\textbackslash \mathrm{n}$ Post: \{TEXT\}$\textbackslash \mathrm{n}$ Possible Relevant Passages: \{PASSAGE\} \\
\midrule
\rowcolor{gray!8}\multicolumn{2}{l}{\textbf{MedXpertQA-Exam, MedQA-Diag, and PMC-Treatment datasets}} \\
\midrule
HyDE & Please write a passage to help answer the Medical Exam.$\textbackslash \mathrm{n}$ Exam: \{TEXT\}$\textbackslash \mathrm{n}$ Passage: \\
\midrule
\multirow{2}{5em}{Query2Doc} & Write a passage to help answer the Medical Exam.$\textbackslash \mathrm{n}$ Examples: \{EXAMPLE\}\\
&Real Test:$\textbackslash \mathrm{n}$ Exam: \{TEXT\}$\textbackslash \mathrm{n}$ Passage: \\
\midrule
\multirow{2}{5em}{LameR} &Give a Medical Exam and its possible relevant passages (most of these passages are wrong). Please write a\\
& correct passage that help answer the exam.$\textbackslash \mathrm{n}$ Exam: \{TEXT\}$\textbackslash \mathrm{n}$ Possible Relevant Passages: \{PASSAGE\}\\
\midrule
\rowcolor{gray!8}\multicolumn{2}{l}{\textbf{PMC-Clinical and IIYi-Clinical datasets}} \\
\midrule
HyDE & Please write a similar case to help diagnose the Clinical Case.$\textbackslash \mathrm{n}$ Case: \{TEXT\}$\textbackslash \mathrm{n}$ Similar Case: \\
\midrule
\multirow{2}{5em}{Query2Doc} &Write a similar case to help diagnose the Clinical Case.$\textbackslash \mathrm{n}$ Examples: \{EXAMPLE\}\\
&Real Test:$\textbackslash \mathrm{n}$ Case: \{TEXT\}$\textbackslash \mathrm{n}$ Similar Case:  \\
\midrule
\multirow{2}{5em}{LameR} &Give a Clinical Case and its possible similar cases (most of these cases are non-similar). Please write a\\
&correct similar case that help diagnose the case.$\textbackslash \mathrm{n}$ Case: \{TEXT\}$\textbackslash \mathrm{n}$ Possible Similar Cases: \{PASSAGE\}\\
\bottomrule
\end{tabular}
}
\label{tab:gar_instruction}
\end{table*}

This section provides comprehensive evaluation results on the R2MED benchmark. Table~\ref{tab:retrieval_result_precision} and Table~\ref{tab:retrieval_result_recall} report the precision@10 and recall@10 scores of 15 retrieval models. We further present the performance of generation-augmented retrieval (GAR) methods in Table~\ref{tab:gar_result_bm25}-\ref{tab:gar_result_nv}, based on three underlying retrievers: BM25, BGE-Large, and NV-Embed-v2. Additionally, Table~\ref{tab:lrm_result_bm25}-\ref{tab:lrm_result_openai} summarize the results of large reasoning models when combined with BM25, BGE-Large, and OpenAI-3-large as the retrieval backends.

\section{Generative Assistance}
We employ AI assistants to refine our paper’s content, and we leverage AI tools to aid in code development.


\begin{table*}[!t]
    \centering
    \caption{An example from Biology dataset.}
    \fontsize{9pt}{11pt}\selectfont
    \begin{tabular}{p{0.98\linewidth}}
    \midrule
        \rowcolor{gray!20}\textbf{Query (ID: 225)} \\
    \midrule
        How does a tree trunk sprout and grow after being cut?\\
        After I cut trees into logs and remove the branches in winter, they start growing. They sprout out and grow completely normal looking stems and leaves and maintain them all summer. The sprouts mostly appear around the cut branches. Sometimes they last all winter and grow for another year. \\How does it find the energy and water necessary to maintain and grow these stems without ground connection and a water source? \\
    \midrule
        \rowcolor{gray!20}
        \textbf{Answer}\\
    \midrule
        This is basically the same that happens after pruning (http://en.wikipedia.org/wiki/Pruning) and involves a basic hormonal regulation mechanism in the plants. \\What happens is that the cut piece of the wood forms a new meristem (http://en.wikipedia.org/wiki/Meristem) which allows the growth of new organs. What’s important is that there is no other growth happening nearby, since that would hormonally inhibit any further growth. This is why such growths happen once you’ve cut the wood, not before (on the healthy stem). This inhibitory effect is known as apical dominance (http://en.wikipedia.org/wiki/Apical\_dominance), which has now been disabled. \\As to where the energy and water comes from, to some extent it is stored within the branches themselves. That’s why you need to dry them before being able to use them in a fire. However, this growth is pretty limited. Further water is probably collected by condensation of water vapour in the air.
        \\
    \midrule
        \rowcolor{gray!20}
        \textbf{Positive document (ID: tree\_grow\_cut/Meristem\_1.txt)}\\
    \midrule
        Primary meristems[edit]\\Apical meristems give rise to the primary plant body and are responsible for primary growth, or an increase in length or height. Apical meristems may differentiate into three kinds of primary meristem:\\Protoderm: lies around the outside of the stem and develops into the epidermis.\\Procambium: lies just inside of the protoderm and develops into primary xylem and primary phloem. It also produces the vascular cambium, and cork cambium, secondary meristems. The cork cambium further differentiates into the phelloderm (to the inside) and the phellem, or cork (to the outside). All three of these layers (cork cambium, phellem, and phelloderm) constitute the periderm. In roots, the procambium can also give rise to the pericycle, which produces lateral roots in eudicots.\\Ground meristem: Composed of parenchyma, collenchyma and sclerenchyma cells that develop into the cortex and the pith.\\
    \midrule
        \rowcolor{gray!20}
        \textbf{Negative document (ID: tree\_grow\_cut/how-to-manage-tree-suckers\_2\_0.txt)}\\
    \midrule
        During the summer, your  trees  will bloom with beautiful leaves in the heat. However, hotter weather usually encourages some tree suckers. To maintain your tree's aesthetic, you'll want to stop these tree suckers from growing. Unfortunately, any tree can suffer from developing the growth of tree suckers, which is likely to happen if the tree is injured or under stress! Not to worry if you've never dealt with these suckers, as stopping tree suckers is simple and effective with the right tools and steps!\\
    \bottomrule
    \end{tabular}
    \label{tab:example_biology}
\end{table*}
\begin{table*}[!t]
    \centering
    \caption{An example from Bioinformatics dataset.}
    \fontsize{9pt}{11pt}\selectfont
    \begin{tabular}{p{0.98\linewidth}}
    \midrule
        \rowcolor{gray!20}\textbf{Query (ID: 21)} \\
    \midrule
        How to select high quality structures from the Protein Data Bank?\\Models of structures deposited in the Protein Data Bank vary in the quality, depending both on the data quality and expertise and patience of the person who built the model. Is there a well-accepted subset of the PDB entries that has only "high quality" structures? 
        
        Ideally these structures would be representative for classes of proteins in the whole PDB. \\
    \midrule
        \rowcolor{gray!20}
        \textbf{Answer}\\
    \midrule
        There is a very nice database, pdbcull (http://dunbrack.fccc.edu/Guoli/pisces\_download.php) (also known as the PISCES server in the literature). It filters the PDB for high resolution and reduced sequence identity. It also seems to be updated regularly. Depending on the cut-offs, you get between 3000 and 35000 structures. \\If you are specifically interested in rotamers, you may want to look at top8000 instead, where they have checked for high resolution, and good MolProbity scores. They also provide a rotamer database. \\PDB also provides their own clustering. They first cluster the sequences, and then extract a representative structure for each one, based on the quality factor (1/resolution - R\_value). This has the advantage of being comprehensive, but you will have bad structures when no good ones were ever obtained.\\
    \midrule
        \rowcolor{gray!20}
        \textbf{Positive document (ID: Pisces:\_A\_Protein\_Sequence\_1)}\\
    \midrule
        For each calculated list, the server provides an output list of accession IDs (e.g., 1ABCA) with sequence length, structure determination method, resolution, and R-factor (if available) and a file of the sequences in FASTA format. The email containing links to these files will be emailed to the user upon completion of the calculation, and will be stored for at least one week.\\ PDB sequences are updated weekly from the PDB mmCIF files.\\ PISCES correctly handles multi-character chain IDs, which are now used in very large structures by the PDB (and some small structures for no good reason).\\ PISCES now allows the user to select whether to include X-ray, NMR, or cryo-EM structures in the output lists.\\
    \midrule
        \rowcolor{gray!20}
        \textbf{Negative document (ID: Amyloid\_13)}\\
    \midrule
    Combined, these methods have provided 3D atomic structures of amyloid fibrils formed by amyloid $\beta$ peptides, $\alpha$-synuclein, tau, and the FUS protein, associated with various neurodegenerative diseases.[46,47]
    X-ray diffraction studies of microcrystals revealed atomistic details of core region of amyloid, although only for simplified peptides having a length remarkably shorter than that of peptides or proteins involved in disease.[48,49]
    The crystallographic structures show that short stretches from amyloid-prone regions of amyloidogenic proteins run perpendicular to the filament axis, consistent with the ``cross-$\beta$'' feature of amyloid structure.\\
    \bottomrule
    \end{tabular}
    \label{tab:example_bioin}
\end{table*}
\begin{table*}[!t]
    \centering
    \caption{An example from Medical Sciences dataset.}
    \fontsize{9pt}{11pt}\selectfont
    \begin{tabular}{p{0.98\linewidth}}
    \midrule
        \rowcolor{gray!20}\textbf{Query (ID: 2)} \\
    \midrule
        Why do corticosteroids harm COVID-19 patients?\\This Science Daily article states that steroids could do more harm than good in treating coronavirus (COVID-19), referring to this The Lancet article.\\Understanding the evidence for harm or benefit from corticosteroids in 2019-nCoV is of immediate clinical importance.\\It's unclear to me how exactly the use of corticosteroids (which is a common treatments for Asthma patients) can do harm other than the vague explanation of \"steroids also impair the immune system's ability to fight viruses\".\\How exactly does the use of steroids harm the body in case of fighting a virus such as SARS-CoV-2 (2019-nCoV) that causes COVID-19?\\
    \midrule
        \rowcolor{gray!20}
        \textbf{Answer}\\
    \midrule
        One of the primary purposes of corticosteroids is to suppress immune activity and inflammation: that's exactly why they are used in asthma.\\Of course the immune system has an actual job besides causing nuisance inflammation: fighting infection.\\For some infections, the harm to the infected person caused by the immune reaction itself is worse than that of the pathogen itself, so steroids can help prevent damage or ease symptoms while the immune system continues to mount a response, without being hyperactive.
        For others, immune suppression may be detrimental.
        \\Corticosteroids mediate the immune response by suppressing a variety of cytokines and increasing others, which influences activity circulating numbers of different immune cell populations. Cortisol binds the glucocorticoid receptor (https://en.wikipedia.org/wiki/Glucocorticoid\_receptor), causing numerous changes in gene expression, many of which are associated with the immune system. From Wikipedia (https://en.wikipedia.org/wiki/Cortisol\#Immune\_response) \\In other words, corticosteroids work by by suppressing the generalized killing/cleanup part of the immune system mediated by macrophages and CD8+ cells, while preserving the B-cell part of the immune system that produces specific antibodies and the neutrophils that phagocytose antibody- and complement-bound pathogens.\\...\\
    \midrule
        \rowcolor{gray!20}
        \textbf{Positive document (ID: Cortisol\_7)}\\
    \midrule
        Immune response[edit] Cortisol prevents the release of substances in the body that cause inflammation. It is used to treat conditions resulting from overactivity of the B-cell-mediated antibody response. Examples include inflammatory and rheumatoid diseases, as well as allergies. Low-dose topical hydrocortisone, available as a nonprescription medicine in some countries, is used to treat skin problems such as rashes and eczema. Cortisol inhibits production of interleukin 12 (IL-12), interferon gamma (IFN-gamma), IFN-alpha, and tumor necrosis factor alpha (TNF-alpha) by antigen-presenting cells (APCs) and T helper cells (Th1 cells), but upregulates interleukin 4, interleukin 10, and interleukin 13 by Th2 cells. This results in a shift toward a Th2 immune response rather than general immunosuppression.\\
    \midrule
        \rowcolor{gray!20}
        \textbf{Negative document (ID: Stress\_Hormone\_6)}\\
    \midrule
        However, the organism makes antibodies against this viral protein, and those antibodies also kill the human ACTH hormone, which leads to the suppression of adrenal gland function. Such adrenal suppression is a way for a virus to evade immune detection and elimination.[15][4][6] This viral strategy can have severe consequences for the host (human that is infected by the virus), as cortisol is essential for regulating various physiological processes, such as metabolism, blood pressure, inflammation, and immune response.\\
    \bottomrule
    \end{tabular}
    \label{tab:example_medical}
\end{table*}
\begin{table*}[!t]
    \centering
    \caption{An example from MedXpertQA-Diag dataset.}
    \fontsize{9pt}{11pt}\selectfont
    \begin{tabular}{p{0.98\linewidth}}
    \midrule
        \rowcolor{gray!20}\textbf{Query (ID: Text-162)} \\
    \midrule
        An 82-year-old woman with a history of moderate aortic stenosis and a drug-eluting stent for coronary artery disease is admitted to the intensive care unit with a one-day history of symptomatic hypotension and melena. She takes aspirin, clopidogrel, simvastatin, metoprolol, and lisinopril. She is resuscitated with intravenous fluids and blood products, is started on high-dose proton-pump-inhibitor therapy, and stabilizes without further bleeding. Upper and lower endoscopy are unremarkable. What test is the most appropriate next step in evaluating this patient? \\
    \midrule
        \rowcolor{gray!20}
        \textbf{Answer}\\
    \midrule
        Video-capsule endoscopy\\
    \midrule
        \rowcolor{gray!20}
        \textbf{Reasoning path}\\
    \midrule
        Let's break down the problem step-by-step:\\1. Patient Profile and Presentation: The patient is an 82‐year‐old woman with a history of moderate aortic stenosis and coronary artery disease (with a drug-eluting stent), currently on dual antiplatelet therapy (aspirin and clopidogrel). She presents with symptomatic hypotension and melena, which are suggestive of gastrointestinal bleeding.\\2. Initial Management: She was resuscitated with intravenous fluids and blood products. High-dose proton-pump inhibitors were started, and her bleeding stabilized. Upper and lower endoscopies were performed but did not reveal any bleeding source.\\3. Diagnostic Challenge: When both upper and lower endoscopies are unremarkable in a patient with suspected GI bleeding, the most likely source is the small bowel. The small bowel is not accessible via standard endoscopy, which is why further evaluation is necessary.\\4. Next Diagnostic Step: The next appropriate test for evaluating small bowel bleeding is video-capsule endoscopy. This non-invasive procedure allows visualization of the entire small intestine and can detect lesions like angioectasia, tumors, or other small-bowel abnormalities that could be causing occult bleeding.\\5. Conclusion:In summary, once upper and lower endoscopies are negative in a patient with GI bleeding, video-capsule endoscopy is the next best step in evaluation.\\Final Answer: Video-capsule endoscopy.\\
    \midrule
        \rowcolor{gray!20}
        \textbf{Positive document (ID: wiki20220301en160\_42780)}\\
    \midrule
        Common reasons for using capsule endoscopy include diagnosis of unexplained bleeding, iron deficiency, or abdominal pain, searching for polyps, ulcers and tumors of small intestine, and diagnosis of inflammatory bowel disease. The images collected by the miniature camera during a session are transferred wirelessly to an external receiver worn by the patient, using any one of a band of appropriate frequencies. The collected images are then transferred to a computer for display, review and diagnosis. A transmitted radio-frequency signal can be used to accurately estimate the location of the capsule and to track it in real time inside the body and gastrointestinal tract. It is unclear if capsule endoscopy can replace gastroscopy for those with cirrhosis.\\
    \midrule
        \rowcolor{gray!20}
        \textbf{Negative document (ID: wiki20220301en372\_3997)}\\
    \midrule
        He co-chairs the International Conference of Capsule Endoscopy and coordinated the Consensus Conference statements to guide capsule usage. He is past president of the New York Society for Gastrointestinal Endoscopy and New York Academy of Gastroenterology and has served on the board of the American College of Gastroenterology. Lewis has written over 72 scientific papers and 27 chapters and he has authored 3 books; all in the field of gastrointestinal endoscopy. He co-authored the technical document behind the American Gastroenterological Association's position statement concerning occult and obscure gastrointestinal bleeding and the ICCE consensus statement for clinical application of capsule endoscopy. He also helped develop a scoring index for inflammatory bowel disease seen on capsule endoscopy.\\
    \bottomrule
    \end{tabular}
    \label{tab:example_medxpert}
\end{table*}
\begin{table*}[!t]
    \centering
    \caption{An example from MedQA-Diag dataset.}
    \fontsize{9pt}{11pt}\selectfont
    \begin{tabular}{p{0.98\linewidth}}
    \midrule
        \rowcolor{gray!20}\textbf{Query (ID: 225)} \\
    \midrule
        A 19-year-old Caucasian male presents to your office with hypopigmented skin. He undergoes a skin biopsy and is found to have an absence of melanocytes in the epidermis. What is the most likely diagnosis? \\
    \midrule
        \rowcolor{gray!20}
        \textbf{Answer}\\
    \midrule
        Vitiligo\\
    \midrule
        \rowcolor{gray!20}
        \textbf{Reasoning path}\\
    \midrule
        Step 1: Identify the key clinical findings.\\ The patient is a 19-year-old Caucasian male. He presents with hypopigmented (depigmented) areas on his skin. A skin biopsy reveals an absence of melanocytes in the epidermis.\\ Step 2: Consider the differential diagnosis for hypopigmentation. \\Vitiligo: Autoimmune destruction of melanocytes leading to depigmented patches. Pityriasis alba: Generally shows mild hypopigmentation with retained melanocytes. Tinea versicolor: A fungal infection that causes discoloration but does not eliminate melanocytes. Nevus depigmentosus: A congenital condition with stable hypopigmentation, but melanocytes are typically present, albeit functionally abnormal. \\Step 3: Evaluate the biopsy finding.\\ An absence of melanocytes is the hallmark of vitiligo. In other conditions causing hypopigmentation, melanocytes may be reduced or dysfunctional, but they are not completely absent. \\Step 4: Arrive at the likely diagnosis. \\Given the clinical presentation and the biopsy evidence (complete absence of melanocytes), the most likely diagnosis is vitiligo. \\Final Answer: Vitiligo.\\
    \midrule
        \rowcolor{gray!20}
        \textbf{Positive document (ID: First\_Aid\_Step2\_154)}\\
    \midrule
        These spots then expand, sometimes in dermatomal patterns, to include large segments of skin. The disease is usually chronic and progressive, with some patients becoming completely depigmented. Many patients have serologic markers of autoimmune disease (e.g., antithyroid antibodies, DM, pernicious anemia) but seldom present with these diseases. Patients with malignant melanoma may develop an antimelanocyte immune response that leads to vitiligo. Diagnosed by the history and clinical picture, with histology demonstrating total absence of melanocytes. Conditions to rule out include postinﬂ ammatory hypopigmentation, scleroderma, piebaldism, and toxic exposure (pheno-lated cleansers are toxic to melanocytes).\\
    \midrule
        \rowcolor{gray!20}
        \textbf{Negative document (ID: InternalMed\_Harrison\_4071)}\\
    \midrule
        disorders, occurring in up to 30\% of patients with vitiligo. Circulating autoantibodies are often found, and the most common ones are antithyroglobulin, antimicrosomal, and antithyroid-stimulating hormone receptor antibodies.\\
    \bottomrule
    \end{tabular}
    \label{tab:example_medqa}
\end{table*}
\begin{table*}[!t]
    \centering
    \caption{An example from PMC-Treatment dataset.}
    \fontsize{9pt}{11pt}\selectfont
    \begin{tabular}{p{0.98\linewidth}}
    \midrule
        \rowcolor{gray!20}\textbf{Query (ID: q\_PMC11551046)} \\
    \midrule
        Case Summary: - Patient Demographics: 44-year-old female - Chief Complaint: Abdominal pain - History of Present Illness: - Presented with abdominal pain in November 2018. - In January 2019, CT scans revealed multiple pelvic masses, peritoneal and omental masses, and metastatic lesions in the liver and lungs. - Carbohydrate antigen 125 (CA125) elevated at 486.0 U/mL. - Diagnosed with stage IVb high-grade serous adenocarcinoma of the ovary in January 2019 via laparoscopic biopsy. - Past Medical History:  - Hysterectomy at age 39 for uterine fibroids.  - No family history of cancer. - Allergies: Not reported. - Family History: No history of cancer. - Social History: Not detailed. - Physical Exam: - Multiple metastatic lesions were documented (liver, lungs, peritoneum, and lymph nodes). - Lab Events: - Elevated CA125 at 486.0 U/mL. - Histopathological results from biopsy confirmed high-grade serous adenocarcinoma. - Imaging Events: - Initial CT scans showed extensive metastatic disease in the pelvis, peritoneum, omentum, liver, and lungs. - Subsequent imaging after chemotherapy showed significant reduction of lesions. - Final Diagnostic Results: Stage IVb high-grade serous adenocarcinoma of ovarian origin. \\Based on the above case summary, what would be an appropriate treatment plan for this patient? \\
    \midrule
        \rowcolor{gray!20}
        \textbf{Answer}\\
    \midrule
        Administer apatinib (500-750 mg orally, daily) as first-line maintenance therapy following chemotherapy and interval debulking surgery.\\
    \midrule
        \rowcolor{gray!20}
        \textbf{Reasoning path}\\
    \midrule
       - Objectives of the Treatment Plan:\\  1. Achieve tumor volume reduction through neoadjuvant chemotherapy and interval debulking surgery.\\  2. Prevent recurrence and extend progression-free survival through maintenance therapy.\\- Rationale:  \\- The patient had advanced-stage ovarian cancer with extensive metastases requiring aggressive primary treatment using neoadjuvant chemotherapy followed by surgery. \\- Adjuvant chemotherapy was employed to further reduce residual disease. \\- Given the high costs of bevacizumab and PARPi, the patient declined these options and opted for the less expensive off-label maintenance therapy using apatinib, which is a VEGFR-2 tyrosine kinase inhibitor. This choice aligns with prior evidence supporting apatinib's anti-angiogenic efficacy in inhibiting tumor-associated vasculature.\\
    \midrule
        \rowcolor{gray!20}
        \textbf{Positive document (ID: PMC11551046\_0)}\\
    \midrule
        Advanced ovarian cancer has a poor prognosis. In recent years, bevacizumab and PARPi have been shown to improve the prognosis of patients with advanced ovarian cancer. Tumor growth requires blood vessels to provide oxygen and nutrients. Furthermore, the vascular endothelial growth factor (VEGF) plays an important role in the process of angiogenesis. Notably, VEGF expression is higher in ovarian tumor tissues than in normal and benign ovarian tissues. There are two different clinical treatment strategies for VEGF. Bevacizumab, an anti-VEGF monoclonal antibody, inhibits the proangiogenic effect of VEGF. Another clinical treatment strategy targeting VEGF is inhibiting the function of the VEGF receptor (VEGFR). VEGF, especially VEGF-A, has been identified as a key factor in tumor angiogenesis. Bevacizumab inhibits the binding of VEGF-A to VEGFR tyrosine kinases (VEGFR1-3), inhibits tumor vascular growth, promotes tumor vascular normalization, and causes tumor cell death. Moreover, VEGFR2 is the main signaling pathway of VEGFR in vascular endothelial cells. Apatinib is a small oral VEGFR-2 tyrosine kinase inhibitor molecule that inhibits tumor angiogenesis by blocking downstream signaling.\\
    \midrule
        \rowcolor{gray!20}
        \textbf{Negative document (ID: PMC11701218\_4)}\\
    \midrule
        Anlotinib is a multi-target tyrosine kinase inhibitor (TKI). It can inhibit various targets including vascular endothelial growth factor receptors (VEGFR), platelet-derived growth factor receptors (PDGFR), fibroblast growth factor receptors (FGFR), and c-Kit (13). It inhibits tumor cell proliferation, invasion, and metastasis and reduces tumor blood supply by inhibiting tumor angiogenesis and growth factor signaling pathways. Compared with other TKIs, anlotinib has better anti-angiogenic activity and higher selectivity, with a significantly lower incidence of grade 3 or higher adverse effects (14, 15).\\
    \bottomrule
    \end{tabular}
    \label{tab:example_pmct}
\end{table*}
\begin{table*}[!t]
    \centering
    \caption{An example from PMC-Clinical dataset.}
    \fontsize{9pt}{11pt}\selectfont
    \begin{tabular}{p{0.98\linewidth}}
    \midrule
        \rowcolor{gray!20}\textbf{Query (ID: q\_PMC3901178)} \\
    \midrule
        A 19-year-old girl was admitted to the gynecology clinic with the complaint of primary amenorrhea. She appears phenotypically female and has normal external genitalia and breast development, but pubic and axillary hair are absent. There are bilateral palpable masses in the inguinal regions. Trans abdominal ultrasonography reveals these masses as immature testes, and a karyotype analysis shows a 46, XY chromosomal pattern. Routine blood tests are normal, but there is slightly elevated luteinizing hormone (LH).\\ What is the most likely diagnosis? \\
    \midrule
        \rowcolor{gray!20}
        \textbf{Answer}\\
    \midrule
        Complete Androgen Insensitivity Syndrome (CAIS).\\
    \midrule
        \rowcolor{gray!20}
        \textbf{Reasoning path}\\
    \midrule
        Let's break down the case step by step:\\1. The patient is a 19‐year-old with primary amenorrhea (no menstruation by age 15–16 is concerning). This sets the stage for a disorder of sexual development.\\2. The patient is phenotypically female with normal external female genitalia and breast development. Normal breasts indicate that estrogen is present, which often results from the aromatization of androgens.\\3. The marked absence of pubic and axillary hair is an important clue. The development of this hair is dependent on androgens. This suggests there is a problem with androgen action despite normal or elevated androgen levels.\\4. Bilateral palpable masses in the inguinal regions are identified on physical exam. These are later confirmed by ultrasonography to be immature testes. Finding testes in a patient with a female phenotype is a significant finding.\\...\\Putting all this together: The patient has a 46,XY karyotype, presence of testes, normal breast development (due to aromatization of androgens into estrogens), absence of pubic/axillary hair (due to androgen receptor insensitivity), and primary amenorrhea because she lacks a uterus. These findings are classic for Complete Androgen Insensitivity Syndrome (CAIS).\\Thus, the most likely diagnosis is Complete Androgen Insensitivity Syndrome (CAIS).\\
    \midrule
        \rowcolor{gray!20}
        \textbf{Positive document (ID: PMC3901178\_2)}\\
    \midrule
        A 22-year-old woman referred to endocrinology and gynecology clinics soon after the operation on her younger sister (Case 1). Her medical history was similar to that of her sister with the symptom of primary amenorrhea. She was recently married and described no sexual problem during intercourse. She had full breast development and feminine appearance of external genitalia with sparse pubic hair. A long and blind ending vagina was found in colposcopy. There were bilateral inguinal mobile masses on palpation that resembled testes on ultrasonography. Neither uterus nor were ovaries demonstrated on the scanning of the abdomen with ultrasonography. Her karyotype was 46, XY and the level of testosterone in peripheral blood was higher than the normal female range. The other biochemical measurements were within normal limits. The patient was diagnosed as CAIS like her 19-year-old sister and her disease was explained to her with the help of a psychologist. Bilateral inguinal orchiectomy was performed in urology clinic and she was discharged at second postoperative day without complication. Histopathologic report of surgical specimen was Sertoli cell adenomas with atrophic seminiferous tubules and Leydig cell hyperplasia.\\...\\
    \midrule
        \rowcolor{gray!20}
        \textbf{Negative document (ID: PMC3850340\_1)}\\
    \midrule
        The proband was an Iranian 17 years-old girl of non-consanguineous parents with a history of irregular menstruation, every 7-8 months. No history of birth defects, mental retardation and congenital malformations was seen in the family. Pregnancy was normal and at birth the patient had no dysmorphic characteristics. In ultrasound examination, her uterine was normal but her left ovarian gonad was small and atrophic without any follicles. Right ovarian gonad was not seen. Her serum follicle-stimulating hormone (FSH) and luteinizing hormone (LH) levels were high, 151 mlU/mL and 56 mlU/mL respectively. Her height was 160 cm at the time of diagnosis. GTG banded karyotype of the patient showed a de novo reciprocal translocations between chromosomes X and 9. Her karyotype was 46, X, rcp (X; 9) (q24; q13)dn.\\
        ...\\
    \bottomrule
    \end{tabular}
    \label{tab:example_pmcc}
\end{table*}
\begin{table*}[!t]
    \centering
    \caption{An example from IIYi-Clinical dataset.}
    \fontsize{9pt}{11pt}\selectfont
    \begin{tabular}{p{0.98\linewidth}}
    \midrule
        \rowcolor{gray!20}\textbf{Query (ID: 27332-1)} \\
    \midrule
        Chief Complaint: A 56-year-old married female farmer presents with a 3-day history of cough and dyspnea, which has worsened over the past day.\\Present Illness: The patient developed dyspnea, cough, and sputum production when the weather turned cool. The sputum is minimal and difficult to expectorate. Initially, the symptoms were not severe but gradually worsened. There is no palpitation, and the symptoms are more pronounced when lying down. She has intermittently taken bronchodilator medication (details of the medication are unknown).\\ Past Medical History: The patient has a 10-year history of asthma, with episodes occurring once or twice almost every year (each episode requiring several days of intravenous therapy). She is otherwise normal. She has no history of smoking or alcohol use. \\Physical Examination: Bilateral wheezing is heard on auscultation of the lungs. Heart sounds are regular at 90 beats per minute. Blood pressure is 130/80 mmHg. Respiratory rate is 20 breaths per minute. \\Ancillary Examinations: None.\\Please infer the possible initial diagnosis based on the above clinical information? \\
    \midrule
        \rowcolor{gray!20}
        \textbf{Answer}\\
    \midrule
        Bronchial Asthma\\
    \midrule
        \rowcolor{gray!20}
        \textbf{Reasoning path}\\
    \midrule
        Below is a systematic analysis and clinical reasoning process for this case: \\
        ...\\Diagnostic Reasoning: The patient has a well-documented history of asthma. The current presentation, characterized by cold air-induced symptoms (cough and dyspnea) and the presence of wheezing on physical examination, strongly supports the diagnosis of bronchial asthma. The absence of other abnormal findings, along with no history of smoking or alcohol use, helps exclude alternative diagnoses such as chronic obstructive pulmonary disease (COPD), further reinforcing the likelihood of an asthma relapse.\\Conclusion: Based on the clinical history, physical findings, and prior diagnosis, a provisional diagnosis of bronchial asthma is established.\\
    \midrule
        \rowcolor{gray!20}
        \textbf{Positive document (ID: 61554-1)}\\
    \midrule
        Case Presentation:  Chief Complaint:  The patient was admitted to the hospital due to \"dyspnea and shortness of breath for over 1 month\",  Present Illness  Over 1 month before admission, the patient developed dyspnea and shortness of breath after exposure to cold air, accompanied by coughing, which was intermittent and non-spastic, with white sticky sputum that was difficult to expectorate, chest tightness, and frequent episodes of awakening from sleep due to breathlessness. There was no chest pain, no fever, no hemoptysis, no nausea, or vomiting. Symptoms worsened after physical activity and improved with rest, being milder during the day and more severe at night. The patient received intravenous treatment at a large hospital and a local hospital, but the details were not provided.\\
        ...\\
        Diagnosis Results  1. Bronchial asthma 2. Hypertension grade 2 (very high-risk group) 3. Arrhythmia Sinus tachycardia. Analysis and Summary: Bronchial asthma often occurs or worsens at night and/or in the early morning, and most patients can spontaneously relieve or be relieved after treatment.\\
    \midrule
        \rowcolor{gray!20}
        \textbf{Negative document (ID: 66438-1)}\\
    \midrule
        Case Presentation: Chief Complaint Recurrent cough for ten years, intermittent cough and wheezing for 2 months. Current Medical History  The patient has had a cough since a cold in winter ten years ago, with little sputum, mainly dry cough. Treatment with Western medicine (specific drugs unknown) was not significantly effective. The symptoms gradually with warmer weather. Since then, the cough has recurred each winter due to colds or exposure to cold, lasting about one month, mostly dry cough, occasionally with shortness of breath, and slight relief was obtained by taking Yin-nourishing and lung-clearing pills and other drugs.\\
        ...\\
        Diagnosis: Cough: Syndrome of Blood Stasis and Lung Involvement Diagnostic Basis The patient has had recurrent coughing for many years, which has damaged the lung qi and lung yin, leading to a deficiency of the body's vital energy. \\
    \bottomrule
    \end{tabular}
    \label{tab:example_iiyi}
\end{table*}
\begin{table*}[b!]
\setlength{\tabcolsep}{3pt}
\caption{The performance of retrieval models on R2MED measured by Precision@10.}
\centering
\resizebox{\textwidth}{!}{
\begin{tabular}{l|c|ccc|ccc|cc|c}
\toprule 
Task &\multirow{2}{*}{\centering Size} & \multicolumn{3}{c|}{Q\&A Reference} & \multicolumn{3}{c|}{Clinical Evidence} & \multicolumn{2}{c|}{Clinical Case} & \multirow{2}{*}{\centering Avg.}\\
\cmidrule(r){1-1} \cmidrule(r){3-5} \cmidrule(r){6-8} \cmidrule(r){9-10}
Model& & Biology & Bioin. & MedS.  & MedE.  & MedD. & PMCT. & PMCC. & IIYiC.  \\
\midrule
\rowcolor{gray!8}\multicolumn{11}{l}{\textbf{Sparse Retrieval}} \\
\midrule
BM25~\cite{robertson2009probabilistic} &-  &7.57 &7.92 &6.02 &0.52 &1.36 &5.33 &5.88 &4.57 &4.90 \\
\midrule
\rowcolor{gray!8}\multicolumn{11}{l}{\textbf{Base Size (< 1B)}} \\
\midrule
Contriever~\cite{izacard2021unsupervised}&110M &4.47 &6.1 &7.61 &0.72 &1.44 &2.87 &4.21 &5.81 &4.15\\
MedCPT\mtag~\cite{jin2023medcpt}&220M &0.87 &7.01 &3.98 &0.52 &0.51 &1.73 &3.25 &3.02 &2.61  \\
InstructOR-L~\cite{su2022one} &335M &7.09 &9.61 &10.46 &1.44 &2.29 &4.13 &2.9 &6.28 &5.53\\
BGE-Large~\cite{xiao2024c} &335M &6.31 &9.74 &10 &1.44 &3.81 &6.07 &4.65 &6.59 &6.08 \\
BMRetriever\mtag~\cite{xu2024bmretriever} &410M &5.05 &10 &8.98 &1.24 &3.31 &6.07 &4.56 &7.75 &5.87 \\
\midrule
\rowcolor{gray!8}\multicolumn{11}{l}{\textbf{Large Size (> 1B)}} \\
\midrule
InstructOR-XL~\cite{su2022one}&1.5B &9.61 &10.78 &10.79 &1.65 &2.03 &3.33 &3.95 &7.44 &6.20 \\
BMRetriever-2B\mtag~\cite{xu2024bmretriever} &2B &8.06 &11.3 &11.59 &3.3 &3.9 &8.2 &7.02 &9.15 &7.82 \\
E5-mistral~\cite{wang2023improving} &7B &8.74 &14.29 &12.5 &2.68 &5.68 &6.2 &8.42 &10.85 &8.67 \\
BMRetriever-7B\mtag~\cite{xu2024bmretriever} &7B &10.1 &14.94 &13.64 &3.71 &7.97 &10.27 &8.16 &10.39 &9.90 \\
SFR-Embedding~\cite{meng2024sfrembedding} &7B &9.13 &15.07 &13.98 &3.92 &8.39 &10.2 &9.91 &11.16 &10.22 \\
GritLM-7B~\cite{muennighoff2024generative} &7B &10.97 &14.94 &13.86 &3.5 &9.07 &9.4 &9.74 &11.01 &10.31  \\
NV-Embed-v2~\cite{lee2024nv} &7B &11.75 &16.36 &14.66 &3.92 &8.22 &10.2 &10.53 &6.28 &10.24 \\
Voyage-3~\cite{voyageemb} &- &11.26 &12.47 &11.82 &3.3 &4.41 &10.07 &7.54 &8.84 &8.71 \\
OpenAI-3-large~\cite{openaiemb} &- &11.26 &13.12 &13.64 &4.64 &7.8 &10.87 &7.72 &7.36 &9.55  \\

\bottomrule
\end{tabular}
}
\label{tab:retrieval_result_precision}
\end{table*}
\begin{table*}[t!]
\setlength{\tabcolsep}{3pt}
\caption{The performance of retrieval models on R2MED measured by Recall@10.}
\centering
\resizebox{\textwidth}{!}{
\begin{tabular}{l|c|ccc|ccc|cc|c}
\toprule 
Task &\multirow{2}{*}{\centering Size} & \multicolumn{3}{c|}{Q\&A Reference} & \multicolumn{3}{c|}{Clinical Evidence} & \multicolumn{2}{c|}{Clinical Case} & \multirow{2}{*}{\centering Avg.}\\
\cmidrule(r){1-1} \cmidrule(r){3-5} \cmidrule(r){6-8} \cmidrule(r){9-10}
Model& & Biology & Bioin. & MedS.  & MedE.  & MedD. & PMCT. & PMCC. & IIYiC.  \\
\midrule
\rowcolor{gray!8}\multicolumn{11}{l}{\textbf{Sparse Retrieval}} \\
\midrule
BM25~\cite{robertson2009probabilistic} &-  &21.85 &36.4 &28.7 &1.35 &4.08 &30.51 &27.92 &14.15 &20.62 \\
\midrule
\rowcolor{gray!8}\multicolumn{11}{l}{\textbf{Base Size (< 1B)}} \\
\midrule
Contriever~\cite{izacard2021unsupervised}&110M &11.47 &25.51 &32.19 &3.13 &3.86 &15.56 &18.71 &18.27 &16.09 \\
MedCPT\mtag~\cite{jin2023medcpt}&220M &2.81 &24.74 &16.84 &1.7 &1.29 &11.1 &15.86 &9.26 &10.45 \\
InstructOR-L~\cite{su2022one} &335M &19.41 &38.81 &46.27 &4.79 &6.86 &21.08 &13.23 &17.13 &20.95 \\
BGE-Large~\cite{xiao2024c} &335M &16.55 &40.08 &44.79 &4.94 &11.15 &32.8 &22.3 &17.11 &23.72 \\
BMRetriever\mtag~\cite{xu2024bmretriever} &410M &13.96 &41.92 &41.88 &5.31 &8.01 &31.32 &23.25 &22.65 &23.54 \\
\midrule
\rowcolor{gray!8}\multicolumn{11}{l}{\textbf{Large Size (> 1B)}} \\
\midrule
InstructOR-XL~\cite{su2022one}&1.5B &26.59 &42.45 &49.36 &5.62 &5.24 &18.83 &19.01 &20.5 &23.45\\
BMRetriever-2B\mtag~\cite{xu2024bmretriever} &2B &23.42 &43.09 &49.92 &12.56 &11.49 &42.21 &34.58 &28.11 &30.67 \\
E5-mistral~\cite{wang2023improving} &7B &21.57 &53.49 &53.12 &10.4 &14.11 &28.68 &41.3 &31.49 &31.77 \\
BMRetriever-7B\mtag~\cite{xu2024bmretriever} &7B &28.39 &60.33 &59.54 &15.41 &20.95 &50.88 &37.57 &30.46 &37.94 \\
SFR-Embedding~\cite{meng2024sfrembedding} &7B &22.46 &55.67 &58.71 &15.03 &21.23 &50.51 &48.76 &31.94 &38.04 \\
GritLM-7B~\cite{muennighoff2024generative} &7B &29.84 &57.39 &58.88 &14.74 &22.86 &47.06 &48.1 &31.8 &38.83 \\
NV-Embed-v2~\cite{lee2024nv} &7B &30.95 &65.35 &63.53 &13.65 &20.18 &49.32 &53.07 &18.32 &39.30 \\
Voyage-3~\cite{voyageemb} &- &29.07 &48.25 &55.19 &14.25 &12.36 &52.04 &38.16 &27.17 &34.56 \\
OpenAI-3-large~\cite{openaiemb} &- &28.59 &51.74 &56.38 &17.11 &19.84 &55.77 &39.91 &22.95 &36.54  \\

\bottomrule
\end{tabular}
}
\label{tab:retrieval_result_recall}
\end{table*}
\begin{table}[t!]
\setlength{\tabcolsep}{3pt}
\caption{Average reranking performance
on R2MED using three classic rerankers:
MonoBERT, BGE-Reranker, and RankLLaMA. We report nDCG@10 for three retrievers, BM25, BGE-Large, and NV-Embed-v2.}
\centering
\resizebox{\textwidth}{!}{
\begin{tabular}{l|c|ccc|ccc|cc|c}
\toprule
Reranker&Top-k & Biology & Bioin. & MedS.  & MedE.  & MedD. & PMCT. & PMCC. & IIYiC. &Avg. \\
\midrule
\rowcolor{gray!8}\multicolumn{11}{l}{\textbf{BM25}} \\
\midrule
None &-  &19.19 &21.55 &19.68 &0.66 &2.55 &23.69 &21.66 &12.02 &15.13 \\
\midrule
MonoBERT&10 &16.12 &23.57 &21.45 &0.93 &3.21 &22.61 &21.25 &11.50 &15.08\\
MonoBERT&100 &10.26 &25.62 &29.69 &1.62 &5.55 &20.42 &17.91 &11.17 &15.28\\
\midrule
BGE-Reranker&10&16.61 &27.26 &21.87 &1.18 &3.67 &23.79 &19.29 &11.21 &15.61 \\
BGE-Reranker&100 &13.28 &26.10 &29.48 &2.66 &7.44 &14.47 &12.55 &12.08 &14.76\\
\midrule
RankLLaMA&10 &17.76 &29.30 &23.88 &1.54 &3.25 &30.17 &22.71 &13.02 &17.70\\
RankLLaMA&100 &15.19 &34.02 &32.94 &3.91 &9.03 &40.13 &25.29 &13.43 &21.74\\
\midrule
\rowcolor{gray!8}\multicolumn{11}{l}{\textbf{BGE-Large}} \\
\midrule
None &-  &12.71 &27.04 &27.76 &4.10 &8.33 &26.45 &15.06 &14.72 &17.02\\
\midrule
MonoBERT&10 &12.73 &27.15 &32.25 &3.33 &8.16 &24.03 &16.38 &13.50 &17.19\\
MonoBERT&100 &8.47 &26.08 &30.40 &1.58 &6.03 &20.36 &17.61 &10.93 &15.18\\
\midrule
BGE-Reranker&10 &13.86 &30.91 &34.81 &4.17 &8.61 &21.45 &14.68 &13.49 &17.75\\
BGE-Reranker&100 &12.56 &28.04 &28.21 &4.26 &6.59 &10.33 &10.87 &10.90 &13.97\\
\midrule
RankLLaMA&10 &13.64 &33.10 &37.00 &4.94 &8.88 &32.92 &18.78 &13.05 &20.29\\
RankLLaMA&100 &13.29 &36.28 &34.85 &7.68 &10.87 &39.67 &25.06 &12.62 &22.54\\
\midrule
\rowcolor{gray!8}\multicolumn{11}{l}{\textbf{NV-Embed-v2}} \\
\midrule
None &-  &27.15 &50.10 &47.81 &10.90 &16.72 &44.05 &39.91 &14.81 &31.43 \\
\midrule
MonoBERT&10 &20.02 &43.13 &40.81 &9.40 &14.84 &33.48 &35.77 &14.30 &26.47\\
MonoBERT&100 &7.43 &27.01 &29.66 &3.03 &7.90 &22.84 &20.49 &11.21 &16.20 \\
\midrule
BGE-Reranker&10 &22.79 &44.32 &43.17 &9.14 &16.25 &26.20 &30.21 &12.31 &25.55 \\
BGE-Reranker&100 &14.23 &28.28 &28.55 &5.05 &8.86 &6.02 &12.16 &9.50 &14.08 \\
\midrule
RankLLaMA&10 &22.55 &49.38 &45.07 &10.56 &17.24 &42.66 &38.36 &13.64 &29.93\\
RankLLaMA&100 &15.77 &36.34 &30.08 &8.01 &13.07 &39.27 &27.58 &12.16 &22.79 \\
\bottomrule
\end{tabular}
}
\label{tab:reranking_result}
\end{table}
\begin{table*}[t!]
\setlength{\tabcolsep}{3pt}
\caption{Average nDCG@10 score of
generation-augmented retrieval (GAR) methods on BM25.}
\centering
\resizebox{\textwidth}{!}{
\begin{tabular}{l|l|ccc|ccc|cc|c}
\toprule
Method &LLM& Biology & Bioin. & MedS.  & MedE.  & MedD. & PMCT. & PMCC. & IIYiC. &Avg. \\
\midrule
BM25 &-  &19.19 &21.55 &19.68 &0.66 &2.55 &23.69 &21.66 &12.02 &15.13 \\
\midrule
HyDE&Qwen-7B &48.49 &29.38 &41.75 &12.17 &19.48 &26.63 &24.48 &8.67 &26.38\\
HyDE&Qwen-72B &54.42 &36.96 &42.12 &16.05 &30.84 &35.95 &29.30 &10.89 &32.07\\
HyDE&GPT-4o &57.34 &43.02 &41.22 &26.55 &49.60 &40.61 &32.80 &10.42 &37.70 \\
\midrule
Query2Doc &Qwen-7B &49.21 &31.05 &38.49 &9.54 &15.27 &27.94 &24.81 &9.54 &25.73\\
Query2Doc &Qwen-72B &57.62 &39.49 &41.64 &15.51 &28.05 &34.53 &31.83 &10.77 &32.43 \\
Query2Doc &GPT-4o &53.32 &40.34 &43.41 &22.03 &40.74 &40.26 &35.57 &11.16 &35.85 \\
\midrule
LameR &Qwen-7B &36.70 &35.18 &39.67 &4.60 &9.40 &34.92 &29.51 &12.55 &25.32\\
LameR &Qwen-72B &51.03 &44.79 &40.75 &12.98 &20.51 &38.78 &31.62 &14.74 &31.90\\
LameR &GPT-4o &49.07 &46.84 &41.55 &9.88 &21.14 &39.46 &33.11 &15.50 &32.07\\
\bottomrule
\end{tabular}
}
\label{tab:gar_result_bm25}
\end{table*}
\begin{table*}[t!]
\setlength{\tabcolsep}{3pt}
\caption{Average nDCG@10 score of
generation-augmented retrieval (GAR) methods on BGE-Large.}
\centering
\resizebox{\textwidth}{!}{
\begin{tabular}{l|l|ccc|ccc|cc|c}
\toprule
Method &LLM& Biology & Bioin. & MedS.  & MedE.  & MedD. & PMCT. & PMCC. & IIYiC. &Avg. \\
\midrule
BGE-Large &-  &12.71 &27.04 &27.76 &4.10 &8.33 &26.45 &15.06 &14.72 &17.02 \\
\midrule
HyDE&Qwen-7B &19.89 &30.18 &40.70 &12.98 &15.05 &41.02 &17.09 &15.07 &24.00 \\
HyDE&Qwen-72B &21.65 &33.96 &41.16 &16.01 &19.82 &44.03 &21.67 &16.80 &26.89\\
HyDE&GPT-4o &22.59 &32.97 &41.29 &19.45 &27.18 &45.43 &23.28 &16.85 &28.63 \\
\midrule
Query2Doc &Qwen-7B &27.70 &34.56 &42.80 &11.65 &14.84 &23.79 &17.49 &14.73 &23.45 \\
Query2Doc &Qwen-72B &34.86 &41.70 &47.44 &15.86 &28.61 &25.85 &24.39 &14.79 &29.19 \\
Query2Doc &GPT-4o&33.23 &40.20 &44.92 &19.14 &37.96 &29.16 &26.31 &14.85 &30.72 \\
\midrule
LameR &Qwen-7B &20.33 &38.37 &41.70 &7.17 &14.86 &44.64 &20.22 &16.21 &25.44 \\
LameR &Qwen-72B &22.60 &37.73 &43.99 &15.32 &20.38 &45.57 &32.75 &20.59 &29.87 \\
LameR &GPT-4o &27.14 &42.24 &44.86 &12.75 &21.65 &42.94 &32.97 &21.40 &30.74 \\
\bottomrule
\end{tabular}
}
\label{tab:gar_result_bge}
\end{table*}
\begin{table*}[t!]
\setlength{\tabcolsep}{3pt}
\caption{Average nDCG@10 score of
generation-augmented retrieval (GAR) methods on NV-Embed-v2.}
\centering
\resizebox{\textwidth}{!}{
\begin{tabular}{l|l|ccc|ccc|cc|c}
\toprule
Method &LLM& Biology & Bioin. & MedS.  & MedE.  & MedD. & PMCT. & PMCC. & IIYiC. &Avg. \\
\midrule
NV-Embed-v2 &-  &27.15 &50.10 &47.81 &10.90 &16.72 &44.05 &39.91 &14.81 &31.43 \\
\midrule
HyDE&Qwen-7B &30.12 &49.95 &49.39 &13.37 &19.49 &42.99 &38.36 &17.86 &32.69\\
HyDE&Qwen-72B &32.35 &54.43 &49.97 &18.83 &24.47 &47.40 &44.88 &23.68 &37.00\\
HyDE&GPT-4o &33.61 &54.15 &50.83 &23.08 &36.09 &47.35 &48.51 &21.30 &39.37\\
\midrule
Query2Doc &Qwen-7B &33.04 &50.62 &48.34 &15.76 &17.44 &40.00 &33.51 &21.41 &32.52 \\
Query2Doc &Qwen-72B &37.09 &52.13 &51.38 &21.08 &30.01 &42.78 &43.81 &24.03 &37.79 \\
Query2Doc &GPT-4o &37.01 &52.35 &51.95 &25.86 &43.30 &47.10 &52.30 &23.37 &41.66 \\
\midrule
LameR &Qwen-7B &34.26 &50.19 &51.73 &9.38 &17.64 &48.96 &44.70 &19.27 &34.52 \\
LameR &Qwen-72B &36.52 &55.48 &52.94 &20.00 &28.64 &50.20 &44.48 &20.25 &38.56 \\
LameR &GPT-4o &39.32 &57.25 &53.84 &14.53 &27.05 &48.86 &44.70 &19.23 &38.10 \\
\bottomrule
\end{tabular}
}
\label{tab:gar_result_nv}
\end{table*}
\begin{table*}[t!]
\setlength{\tabcolsep}{3pt}
\caption{The nDCG@10 performance of large reasoning models on BM25.}
\centering
\resizebox{\textwidth}{!}{
\begin{tabular}{l|ccc|ccc|cc|c}
\toprule 
Model&  Biology & Bioin. & MedS.  & MedE.  & MedD. & PMCT. & PMCC. & IIYiC. &Avg.  \\
\midrule
BM25  &19.19 &21.55 &19.68 &0.66 &2.55 &23.69 &21.66 &12.02 &15.13  \\
\midrule
\multicolumn{10}{l}{\textbf{Large Language Model}} \\
\midrule
\rowcolor{green!10} Qwen2.5-7B-Ins. &48.49 &29.38 &41.75 &12.17 &19.48 &26.63 &24.48 &8.67 &26.38\\
\rowcolor{gray!10} Qwen2.5-32B-Ins. &52.88 &39.2 &42.42 &16.8 &25.87 &33.27 &26.72 &11.81 &31.12 \\
\rowcolor{blue!10} Llama3.1-70B-Ins. &52.54 &39.42 &41.05 &16.99 &33.87 &37.32 &28.67 &9.32 &32.40 \\
GPT-4o &57.34 &43.02 &41.22 &26.55 &49.6 &40.61 &32.8 &10.42 &37.70 \\
\midrule
\multicolumn{10}{l}{\textbf{Large Reasoning Model}} \\
\midrule
\rowcolor{gray!12} R1-Distill-Qwen-32B &48.9 &38.8 &38.28 &16.04 &25.21 &31.55 &22.47 &11.15 &29.05 \\
\rowcolor{gray!15} QwQ-32B &58.24 &42.35 &42.2 &23.7 &38.12 &41.65 &26.66 &7.34 &35.03 \\
\rowcolor{blue!12} R1-Distill-Llama-70B &49.52 &38.86 &39.48 &22.12 &38.4 &34.95 &33.86 &9.13 &33.29 \\
\rowcolor{blue!17} HuatuoGPT-o1-70B &49.77 &40.02 &38.47 &21.91 &39.54 &39.3 &28.38 &10.01 &33.43 \\
o3-mini &59.65 &46.56 &47.17 &34.64 &55.22 &41.65 &35.32 &7.86 &41.01 \\
\midrule
\multicolumn{10}{l}{\textbf{Search-Enhanced Large Reasoning Model}} \\
\midrule
Search-R1-3B &34.38 &20.31 &30.04 &5.47 &13.18 &31.66 &19.69 &5.53 &20.03 \\
\rowcolor{green!15} Search-R1-7B &36.51 &26.92 &34.62 &10.02 &13.88 &34.98 &28.18 &11.37 &24.56 \\
\rowcolor{gray!20} Search-o1_{\textbf{QwQ-32B}} &60.78 &41.58 &43.93 &24.07 &39.94 &37.52 &28.38 &8.34 &35.57 \\
Search-o1_{\textbf{Qwen3-32B}} &56.84 &39.48 &43.01 &22 &34.42 &41.62 &26.2 &7.16 &33.84 \\
\bottomrule
\end{tabular}
}
\label{tab:lrm_result_bm25}
\end{table*}
\begin{table*}[t!]
\setlength{\tabcolsep}{3pt}
\caption{The nDCG@10 performance of large reasoning models on BGE-Large.}
\centering
\resizebox{\textwidth}{!}{
\begin{tabular}{l|ccc|ccc|cc|c}
\toprule 
Model&  Biology & Bioin. & MedS.  & MedE.  & MedD. & PMCT. & PMCC. & IIYiC. &Avg.  \\
\midrule
BGE-Large  &12.71 &27.04 &27.76 &4.1 &8.33 &26.45 &15.06 &14.72 &17.02 \\
\midrule
\multicolumn{10}{l}{\textbf{Large Language Model}} \\
\midrule
\rowcolor{green!10} Qwen2.5-7B-Ins. &19.89 &30.18 &40.7 &12.98 &15.05 &41.02 &17.09 &15.07 &24.00 \\
\rowcolor{gray!10} Qwen2.5-32B-Ins. &23.6 &33.76 &41.51 &14.41 &17.98 &39.77 &22.23 &16.22 &26.19 \\
\rowcolor{blue!10} Llama3.1-70B-Ins. &21.4 &37.06 &41.44 &15.91 &22.85 &45.78 &24.61 &16.36 &28.18 \\
GPT-4o &22.59 &32.97 &41.29 &19.45 &27.18 &45.43 &23.28 &16.85 &28.63 \\
\midrule
\multicolumn{10}{l}{\textbf{Large Reasoning Model}} \\
\midrule
\rowcolor{gray!12} R1-Distill-Qwen-32B &19.61 &33.42 &40.76 &14.78 &19.3 &42.38 &23.84 &17.1 &26.40  \\
\rowcolor{gray!15} QwQ-32B &21.01 &32.89 &39.45 &16.68 &22.55 &43.4 &20.75 &15.95 &26.59 \\
\rowcolor{blue!12} R1-Distill-Llama-70B &21.68 &34.79 &40.33 &19.06 &23.93 &42.74 &26.44 &18.28 &28.41 \\
\rowcolor{blue!17} HuatuoGPT-o1-70B &18.88 &35.5 &39.72 &22 &27.48 &43.29 &26.46 &18.5 &28.98 \\
o3-mini &22.18 &37.62 &43.97 &25.39 &34.91 &44.55 &24.65 &17.04 &31.29  \\
\midrule
\multicolumn{10}{l}{\textbf{Search-Enhanced Large Reasoning Model}} \\
\midrule
Search-R1-3B &16.8 &26.83 &36.74 &8.4 &12.17 &37.15 &22.84 &16.16 &22.14 \\
\rowcolor{green!15} Search-R1-7B &19.14 &30.38 &37.31 &12.53 &14.81 &38.93 &24.51 &20.6 &24.78 \\
\rowcolor{gray!20} Search-o1_{\textbf{QwQ-32B}} &21.92 &32.34 &42.71 &20.1 &24.34 &42.91 &26.31 &18.66 &28.66 \\
Search-o1_{\textbf{Qwen3-32B}} &23.88 &30.34 &42.16 &18.48 &25.49 &42.96 &25.46 &17.47 &28.28 \\
\bottomrule
\end{tabular}
}
\label{tab:lrm_result_bge}
\end{table*}
\begin{table*}[t!]
\setlength{\tabcolsep}{3pt}
\caption{The nDCG@10 performance of large reasoning models on OpenAI-3-large.}
\centering
\resizebox{\textwidth}{!}{
\begin{tabular}{l|ccc|ccc|cc|c}
\toprule 
Model&  Biology & Bioin. & MedS.  & MedE.  & MedD. & PMCT. & PMCC. & IIYiC. &Avg.  \\
\midrule
OpenAI-3-large  &23.82 &40.51 &44.05 &11.78 &15.01 &47.43 &28.87 &17.12 &28.57 \\
\midrule
\multicolumn{10}{l}{\textbf{Large Language Model}} \\
\midrule
\rowcolor{green!10} Qwen2.5-7B-Ins. &30.15 &42.33 &45.79 &15.45 &19.73 &48.64 &30.39 &19.79 &31.53 \\
\rowcolor{gray!10} Qwen2.5-32B-Ins. &31.37 &45.46 &46.44 &21.13 &24.69 &48.05 &34.45 &22.38 &34.25 \\
\rowcolor{blue!10} Llama3.1-70B-Ins. &31.31 &46.83 &48.14 &21.42 &28.32 &51.22 &37.11 &21.7 &35.76  \\
GPT-4o &32.15 &45.99 &47.97 &27.28 &36.92 &51.24 &38.96 &21.82 &37.79  \\
\midrule
\multicolumn{10}{l}{\textbf{Large Reasoning Model}} \\
\midrule
\rowcolor{gray!12} R1-Distill-Qwen-32B &30.39 &44.89 &47.62 &20.33 &27.89 &49.04 &36.26 &21.04 &34.68  \\
\rowcolor{gray!15} QwQ-32B &32.51 &45.46 &46.76 &24.81 &31.76 &52.78 &35.1 &22.89 &36.51 \\
\rowcolor{blue!12} R1-Distill-Llama-70B &30.08 &47.44 &48.98 &24.94 &33.24 &50.49 &41.39 &22.32 &37.36  \\
\rowcolor{blue!17} HuatuoGPT-o1-70B &29.77 &47.7 &48.4 &27.46 &36.79 &52.93 &39.26 &23.63 &38.24 \\
o3-mini &31.5 &48.61 &48.19 &31.74 &39.51 &51.69 &38.92 &22.56 &39.09 \\
\midrule
\multicolumn{10}{l}{\textbf{Search-Enhanced Large Reasoning Model}} \\
\midrule
Search-R1-3B &24.48 &40.91 &47.65 &11.62 &16.78 &47.79 &32.27 &20 &30.19 \\
\rowcolor{green!15} Search-R1-7B &28.3 &43.6 &48.03 &15.2 &18.57 &48.38 &38.15 &22.92 &32.89  \\
\rowcolor{gray!20} Search-o1_{\textbf{QwQ-32B}} &32.43 &46.95 &47.84 &25.19 &31.22 &51.67 &36.58 &26.51 &37.30\\
Search-o1_{\textbf{Qwen3-32B}} &34.73 &45.38 &47.18 &26.95 &33.71 &50.78 &40.02 &24.17 &37.87 \\
\bottomrule
\end{tabular}
}
\label{tab:lrm_result_openai}
\end{table*}

\end{document}